\documentclass[10pt,journal,doublecolumn,twoside]{IEEEtran}
\usepackage{amsfonts}
\IEEEoverridecommandlockouts

\usepackage{subfigure}
\usepackage{colortbl}
\usepackage{bm}

\usepackage{graphicx}  
\usepackage{url}       

\usepackage{amsmath}   
\usepackage{cite}

\usepackage{amsfonts,amssymb}

\usepackage{stfloats}

\usepackage{cases}
\usepackage{algorithm}
\usepackage{multirow}
\usepackage{algorithmic}
\usepackage{amsthm}
\usepackage[utf8]{inputenc}

\newcommand{\ls}[1]
    {\dimen0=\fontdimen6\the\font
     \lineskip=#1\dimen0
     \advance\lineskip.5\fontdimen5\the\font
     \advance\lineskip-\dimen0
     \lineskiplimit=.9\lineskip
     \baselineskip=\lineskip
     \advance\baselineskip\dimen0
     \normallineskip\lineskip
     \normallineskiplimit\lineskiplimit
     \normalbaselineskip\baselineskip
     \ignorespaces
    }

\newcounter{TempEqCnt}

\usepackage[absolute,showboxes]{textpos}
\TPMargin{3pt}

\begin{document}
\setlength{\columnsep}{0.24in}
\newcommand{\copyrightstatement}{
    \begin{textblock}{0.84}(0.08,0.95) 
         \noindent
         \footnotesize
         \copyright 2018 IEEE. Personal use of this material is permitted. Permission from IEEE must be obtained for all other uses, in any current or future media, including reprinting/republishing this material for advertising or promotional purposes, creating new collective works, for resale or redistribution to servers or lists, or reuse of any copyrighted component of this work in other works. DOI: 10.1109/TVT.2018.2825539.
    \end{textblock}
}
\copyrightstatement
\setlength{\columnsep}{0.24in}

\title{Mode Hopping for Anti-Jamming in Radio Vortex Wireless Communications}
\author{Liping Liang, Wenchi Cheng,~\IEEEmembership{Member,~IEEE}, Wei Zhang,~\IEEEmembership{Fellow,~IEEE}, and Hailin Zhang,~\IEEEmembership{Member,~IEEE}


\thanks{\ls{.5}Copyright (c) 2015 IEEE. Personal use of this material is permitted. However, permission to use this material for any other purposes must be obtained from the IEEE by sending a request to pubspermissions@ieee.org.

This work was supported in part by the National Natural Science Foundation of China (Nos. 61771368 and 61671347), Young Elite Scientists Sponsorship Program By CAST (2016QNRC001), the 111 Project of China (B08038), and the Australian Research Council's Projects funding scheme under Projects (DP160104903 and LP160100672). Part of this work has been accepted by the IEEE/CIC International Conference on Communications in China (ICCC), Qingdao, China, 2017~\cite{ICCC2017_OAM_hop}. 

Liping Liang, Wenchi Cheng, and Hailin Zhang are with the State Key Laboratory of Integrated Services Networks, Xidian University, Xi'an, 710071, China (e-mails: lpliang@stu.xidian.edu.cn; wccheng@xidian.edu.cn (corresponding author); hlzhang@xidian.edu.cn).

Wei Zhang is with the University of New South Wales, Sydney, Australia (e-mail: wzhang@ee.unsw.edu.au).
}
}

\maketitle

\begin{abstract}
Frequency hopping (FH) has been widely used as a powerful technique for anti-jamming in wireless communications. However, as the wireless spectrum becoming more and more crowded, it is very difficult to achieve efficient anti-jamming results with FH based schemes. Orbital angular momentum (OAM), which provides the new angular/mode dimension for wireless communications, offers an intriguing way for anti-jamming. In this paper, we propose to use the orthogonality of OAM-modes for anti-jamming in wireless communications. In particular, we propose the mode hopping (MH) scheme for anti-jamming within the narrow frequency band. We derive the closed-form expression of bit error rate (BER) for multiple users scenario with our developed MH scheme. Our developed MH scheme can achieve the same anti-jamming results within the narrow frequency band as compared with the conventional wideband FH scheme. Furthermore, we propose mode-frequency hopping (MFH) scheme, which jointly uses our developed MH scheme and the conventional FH scheme to further decrease the BER for wireless communication. Numerical results are presented to show that the BER of our developed MH scheme within the narrow frequency band is the same with that of the conventional wideband FH scheme. Moreover, the BER of our developed MFH schemes is much smaller than that of the conventional FH schemes for wireless communications.

\end{abstract}




\section{Introduction}
\IEEEPARstart{F}{r}EQUENCY hopping (FH), which is a solid anti-jamming technique, has been extensively used in wireless communications. There exist some typical FH schemes such as adaptive FH~\cite{AFH_2006}, differential FH~\cite{DFH_2006}, uncoordinated FH~\cite{UFH_2008}, adaptive uncoordinated FH~\cite{UFH_2012}, and message-driven FH~\cite{MDFH_2013}, etc. These FH schemes can achieve efficient anti-jamming results for various wireless communications scenarios.

However, as the wireless spectrum becoming more and more crowded, it is very difficult for FH schemes to satisfy the reliability requirements of wireless communications. The authors of~\cite{Interference_band} pointed out that if the interference covers the whole frequency band, it is very hard to guarantee the reliability of wireless communications with FH.
Also, when the number of channels with partial-band-noise jamming increases to a certain proportion of the total number of channels, the bit
error rate (BER) of FH based wireless communications is relatively large~\cite{2009_PBNJ,2011_PBNJ}.
When the available number of hopping channel is small, the probability jammed by interfering users becomes very high, thus severely downgrading the spectrum efficiency of wireless communications\cite{Efficacy_FH}.
With these limitations in mind, it is highly demanded to achieve highly efficient anti-jamming results for future wireless communications. How to guarantee the reliability of wireless communications still is an open challenge~\cite{2017_Bigdata,2017_Adhoc}.

Recently, more and more academic researchers show their interests in orbital angular momentum (OAM), which is another important property of electromagnetic waves and a result of signal possessing helical phase fronts. There are some studies on OAM in radio wireless communications~\cite{2007_Thide,2013_uncompressed,2013_radio,2018_Mag,2015_radio,Mode_2015,2016_radio,2018_Tcom}. The authors of~\cite{2007_Thide} performed the first OAM experiment in the low frequency domain and showed that OAM is not restricted to the very high frequency range. Then, the authors of \cite{2013_uncompressed,2013_radio,2015_radio,Mode_2015,2016_radio} started to apply OAM based transmission in radio wireless communications. Also, the OAM-based vortex beams can be generated by designing coding metasurfaces based on the Pancharatnam-Berry phase~\cite{metasurface_gernerate} and combined with orthogonal polarizations to encode information of the matesurfaces at the transmitter, thus reducing the information loss~\cite{metasurfaces_app}. In addition, the authors proposed the radio system based OAM~\cite{2015_capacity}, vortex multiple-input-multiple-output (MIMO) communication system~\cite{2015_cap_Access}, and OAM-embedded-MIMO system~\cite{2017_cap} for achieving higher capacity without increasing the bandwidth. The OAM-based vortex waves with different topological charges are mutually orthogonal when they propagate along the same spatial axis~\cite{2012_sameaxis,2013_sameaxis}, and thus can carry a number of independent data streams within a narrow frequency band. Also, we have studied the OAM-mode modulation and OAM waves converging~\cite{GC17_OAM,ICCC2017_OAM_con}.

From the generic viewpoint, OAM can be considered as line of sight (LOS) MIMO because there are multiple antennas/array-elements at the transmitter and receiver. However, there are some differences between OAM and MIMO. As compared with MIMO in the conventional space domain, OAM provides a new mode domain. OAM is the mode multiplexing technique in the field of vortex-electromagnetic beams while MIMO is the spatial multiplexing technique in the field of plane-electromagnetic beams~\cite{2017_MIMO}. Mode multiplexing utilizes the orthogonality of OAM beams to minimize interchannel crosstalk and recover different data streams, thereby avoiding the use of MIMO processing. However, each data stream is received by multiple spatially separated receivers using MIMO spatial multiplexing technique. As compared with the conventional FH schemes in frequency domain, OAM, which provides the angular/mode dimension for wireless communication, offers a new way to achieve efficient anti-jamming results for wireless communications.



\begin{figure}
\centering
\subfigure[Transmitter]
{\label{fig:subfig:a} 
\includegraphics[width=0.9\linewidth]{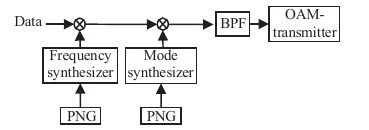}}
\vfill
\subfigure[Receiver]
{\label{fig:subfig:b} 
\includegraphics[width=1\linewidth]{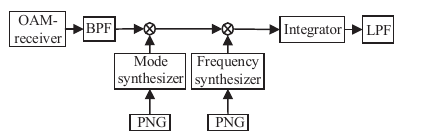}}
\caption{The MFH system model.}
\label{fig:system} 
\vspace{-5pt}
\end{figure}

The authors of~\cite{Gibson_2004,2009_millimeter,2014_Bai,2017_MD} showed that OAM has the potential anti-jamming for wireless communications. In the experiment~\cite{Gibson_2004}, any attempt to sample the OAM based vortex wave will be subject to an angular restriction and a lateral offset, both of which result in inherent uncertainty in the measurement. Thus, the information encoded with OAM-modes is resistant to eavesdropping. Using the OAM to encode the data offers an inherent security enhancement for OAM-based millimeter wave wireless communications~\cite{2009_millimeter}. Moreover, the OAM-mode division multiplexing technique can provide high security for wireless communications~\cite{2017_MD}.
However, how to hop among different OAM-modes for anti-jamming is still an open and challenging problem.

To achieve efficient hopping performance for wireless communications, in this paper we utilize the orthogonality of OAM-modes for anti-jamming. First, we propose the mode hopping (MH) scheme for anti-jamming within the narrow frequency band. We derive the closed-form expression of BER for multiple users scenario with our developed MH scheme. Our developed MH schemes can achieve the same anti-jamming results within the narrow frequency band as compared with the conventional wideband FH scheme. Furthermore, we propose mode-frequency hopping (MFH) scheme, which jointly uses our developed MH and the conventional FH scheme, for better anti-jamming results.  We also derive the closed-form expression of BER for multiple users scenario with our developed MFH scheme, which can further significantly decrease the BER of wireless communications. We conduct extensive numerical results to evaluate our developed schemes, showing that our proposed MH and MFH schemes are superior than the conventional FH schemes.


The rest of this paper is organized as follows. Section II gives the MH and MFH system models. Section III de-hops OAM-mode, decomposes OAM-mode, and derives the closed-form expression of BER for multiple users scenario with our developed MH scheme. Based on the MH scheme, Section IV derives the closed-form expression of BER for multiple users scenario with our developed MFH scheme. Section V evaluates our developed MH and MFH schemes, and compares the BERs of our developed schemes with that of the conventional FH scheme. The paper concludes with Section VI.

\section{MFH System Model}\label{sec:DPSK_sys}
In this section, we build up the MFH system model (including the MH system), an example of which is shown in Fig.~\ref{fig:system}. The MH system consists of OAM-transmitter, mode synthesizer, pseudorandom noise sequence generator (PNG), band pass filter (BPF), OAM-receiver, integrator, and low pass filter (LPF). The MFH system adds two frequency synthesizers based on the MH system.
The OAM-transmitter and OAM-receiver can be uniform circular array (UCA) antenna, which consists of $N$ array-elements distributed equidistantly around the perimeter of circle~\cite{2011_Eletromagnetic}. For the OAM-transmitter, the $N$ array-elements are fed with the same input signal, but with a successive delay from array-element to array-element such that after a full turn the phase has been incremented by an integer multiple $l$ of $2\pi$, where $l$ is the OAM-mode and satisfies with $-N/2 < l \leq N/2$. 
We have $K$ interfering users which may use the same OAM-modes with the desired user, thus causing interference on the desired user.

In our proposed system, one data symbol experiences $U$ OAM-mode hops or frequency hops. At the transmitter, the mode/frequency synthesizer, which is controlled by PNG, selects an OAM-mode or a range of frequency band. To de-hop OAM-mode at the receiver, PNGs are identical to those used in the transmitter. The integrator and low pass filter are used at the receiver to recover the transmit signal. 



\subsection{MH Pattern}

An example of MH pattern is illustrated in Fig.~\ref{fig:subfig:a}, where the OAM-mode resource is divided into $N$ OAM-modes and the time resource is divided into $U$ time-slots. For the MH system, we denote by $t_{h}$ the duration of one time slot, which is also called hop duration. We integrate OAM-mode and time into a two-dimension time-mode resource block. Each hop corresponds to a time-mode resource block. For the $u$-th hop, we denote by $l_{u}$ ($1 \leq u \leq U,  -N/2 < l_{u} \leq N/2)$ the corresponding OAM-mode.

For comparison purpose, we also plot an example of FH pattern in Fig.~\ref{fig:subfig:b}, where the frequency resource is divided into $Q$ frequency bands and the time resource is divided into $U$ time-slots. As shown in Fig.~\ref{fig:subfig:b}, the frequency and time are integrated to form a two-dimension time-frequency resource blocks. Each hop corresponds to a time-frequency resource block. We denote by $q$ ($1\leq q \leq Q$) the index of frequency band. For the $q$-th frequency band, we denote by $F_{q}$ the corresponding carrier frequency. For the $u$-th hop, we denote by $f_{u}$ ($f_{u} \in \{F_{1}, \ldots, F_{q}, \ldots, F_{Q}\} $) the corresponding carrier frequency.


%

\begin{figure}
\centering
\subfigure[MH pattern]
{\label{fig:subfig:a} 
\includegraphics[width=1\linewidth]{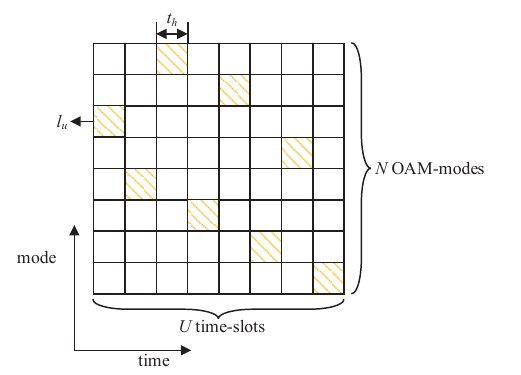}}
\vfill
\subfigure[FH pattern]
{\label{fig:subfig:b} 
\includegraphics[width=1\linewidth]{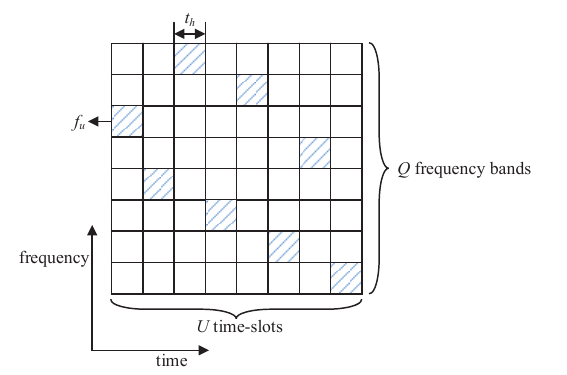}}
\vfill
\subfigure[MFH pattern]
{\label{fig:subfig:c}
\includegraphics[width=1.05\linewidth]{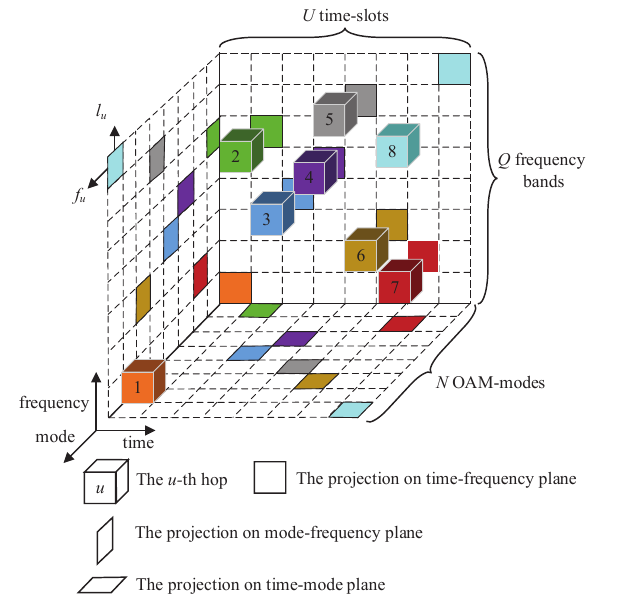}}
\caption{The MH, FH, and MFH patterns.}
\label{fig:subfig} 
\end{figure}

\subsection{MFH Pattern}

An example of MFH pattern is illustrated in Fig.~\ref{fig:subfig:c},  where the OAM-mode resource is divided into $N$ OAM-modes, the frequency resource is divided into $Q$ frequency bands, and the time resource is divided into $U$ time-slots. A cube denotes a hop with respect to the carrier frequency, OAM-mode, and time-slot. Each hop is identified by the specified color. For the $u$-th hop, the corresponding frequency band and OAM-mode are $f_{u}$ ($f_{u} \in \{F_{1}, \ldots, F_{q}, \ldots, F_{Q}\} $) and $l_{u}$ ($-N/2 < l_{u} \leq N/2 $), respectively. We denote by $u$ the index of hop.

As shown in Figs.~\ref{fig:subfig:a} and~\ref{fig:subfig:c}, one data symbol carrying different OAM-modes can be transmitted within $U$ hops. For each hop, the corresponding OAM-mode is one of $N$ OAM-modes controlled by PNG. Also, any interference to impact the desired OAM signal should be with the  same azimuthal angle. However, being with the same azimuthal angle is a small probability event. Thus, the desired OAM signal is resistant to jamming caused by interfering users.


When the available frequency band is relatively narrow, the conventional FH scheme cannot be efficiently used. However, our developed MH scheme can solve the problem mentioned above for anti-jamming without increasing the frequency bands. Using our developed MH scheme, signal can be transmitted using the new mode dimension within the narrow frequency band, thus achieving efficient anti-jamming results in wireless communications. On the other hand, when frequency band is relatively wide, signal can be transmitted within the angular domain and the frequency domain simultaneously by using the MFH scheme, which jointly uses the MH scheme and the FH scheme, to further achieve better anti-jamming results than the conventional FH scheme. In addition, signal can be encoded with orthogonal polarized bit and OAM mode bit~\cite{metasurfaces_app}. Our developed MH and MFH scheme can add orthogonal polarization parameter to achieve better anti-jamming results.

%


In the following, we first study the MH system within the narrow frequency band. Then, we investigate the MFH system by jointly using our developed MH scheme and the conventional FH scheme.

\section{MH Scheme}\label{sec:MH}
In this section, we propose the MH scheme and derive the corresponding BER for our developed MH scheme. First, we give the transmit signal of MH scheme and derive the channel amplitude gain for UCA antenna based tranceiver. Then, at the receiver we de-hop and decompose OAM mode, to recover the transmit signal. Finally, we derive the BER of our developed MH scheme to analyze the anti-jamming performance for MH communications.
\subsection{Transmit Signal}
For MH communications, we denote by $t$ the time variable. The transmit signal, denoted by $x_{1}(u,t)$, for the desired user corresponding to the $u$-th hop can be expressed as follows:
\begin{eqnarray}
   x_{1}(u,t)=s(t)\varepsilon_{u}(t - u t_{h})e^{j \varphi l_{u}},
\end{eqnarray}
where $s(t)$ represents the transmit signal within a symbol duration, $\varphi$ is the azimuthal angle for all users, and $\varepsilon_{u}(t)$ is the rectangular pulse function given by
\begin{eqnarray}
\varepsilon_{u}(t)=\begin{cases}
1 \ \ \ \ \ \ {\rm{if}} \ (u-1)t_{h} \leq t < u t_{h}; \\
0 \ \ \ \ \ \ \rm{otherwise}.
\end{cases}
\end{eqnarray}

We denote by $h_{l_{u}}$ the channel gain corresponding to the channel from the desired OAM-transmitter to the OAM-receiver for the $u$-th hop, $h_{l_{u},k}$ the channel gain corresponding to the channel from the $k$-th ($1\leq k \leq K$) interfering user's OAM-transmitter to the OAM-receiver for the $u$-th hop, $s_{k}(t)$ the transmit signal for the $k$-th interfering user within a symbol duration, $l_{u,k}$ ($-N/2 < l_{u,k} \leq N/2$) the hopping OAM-mode for the $k$-th interfering user corresponding to the $u$-th hop, and $n(u,t)$ the received noise for the $u$-th hop. Then, the received signal, denoted by $r_{1}(u,t)$, for the desired user corresponding to the $u$-th hop can be obtained as follows:
\setcounter{TempEqCnt}{\value{equation}} 
\setcounter{equation}{9} 
\begin{figure*}[ht]
\begin{eqnarray}
r_{1}^{\prime}(u,t)&=&\frac{1}{2\pi} \int_{0}^{2 \pi}\tilde{r}_{1}(u,t) \left(e^{j 2\varphi l_{u}}\right)^{*} d \varphi \nonumber \\
&=&\left\{
\begin{array}{lll}
h_{l_{u}}s(t)\varepsilon(t-u t_{h})+\tilde{n}(u,t),\ \ \hspace{3.5cm} \ \ \ \ \  l_{u,k}\neq l_{u};
\\
h_{l_{u}}s(t)\varepsilon(t-u t_{h})+\sum\limits_{k=1}^{D_{u}}h_{l_{u},k}s_{k}(t)\varepsilon(t-u t_{h})+\tilde{n}(u,t),\ \ \ l_{u,k}= l_{u}.
\\
\end{array}
\right.
\label{eq:M_integrator}
\end{eqnarray}
\hrulefill
\vspace{-10pt}
\end{figure*}
\setcounter{equation}{\value{TempEqCnt}}
\begin{eqnarray}
    r_{1}(u,t)&=&h_{l_{u}}x_{1}(u,t)+n(u,t)\nonumber\\
   &&+\sum_{k=1}^{K}h_{l_{u,k}}s_{k}(t)\varepsilon(t-u t_{h}) e^{j \varphi l_{u,k}}.\label{eq:r_MH}
\end{eqnarray}


For UCA antenna based transceiver,
the pathloss, denoted by $h_{d}$, can be given as follows~\cite{Is_OAM}:
\begin{eqnarray}
    h_{d}=\beta \frac{\lambda}{4 \pi \mid\vec{d} - \vec{r}_{n} \mid}e^{-j \frac{2 \pi \mid\vec{d} - \vec{r}_{n} \mid}{\lambda}},
\end{eqnarray}
where $\vec{d}$ denotes the position vector from the OAM-transmitter to the OAM-receiver in free space, $\vec{r}_{n}$ denotes the position vector from the $n$-th ($1\leq n \leq N$) array-element of the OAM-transmitter to the center of OAM-transmitter, $\beta$ contains all relevant constants such as attenuation and phase rotation caused by antennas and their patterns on both sides, and $\lambda$ is the wave length of carrier.
Thus, the channel amplitude gain, denoted by $h_{l}$, from the OAM-transmitter to the OAM-receiver for the $l$-th OAM-mode can be derived as follows~\cite{Is_OAM}:
\begin{eqnarray}
    h_{l}\hspace{-0.2cm}&=&\hspace{-0.2cm}\sum_{n=1}^{N}\beta \frac{\lambda}{4 \pi \mid\vec{d} - \vec{r}_{n}\mid}e^{-j \frac{2 \pi \mid\vec{d} - \vec{r_{n}}\mid}{\lambda}} e^{j \frac{2 \pi (n-1)}{N} l}\nonumber\\
    \hspace{-0.2cm}&=&\hspace{-0.2cm}\sum_{n=1}^{N}\beta \frac{\lambda}{4 \pi d}e^{-j \frac{2 \pi d}{\lambda}} e^{j\frac{2 \pi \mid\vec{d}\cdot\vec{r}_{n}\mid }{\lambda}}e^{j \frac{2 \pi (n-1)}{N} l}\nonumber\\
    \hspace{-0.2cm}&=&\hspace{-0.2cm}\beta \frac{\lambda}{4 \pi d}e^{-j \frac{2 \pi d}{\lambda}} \sum_{n=1}^{N}e^{j\frac{2 \pi \mid\vec{d}\cdot\vec{r}_{n}\mid }{\lambda}}e^{j \frac{2 \pi (n-1)}{N} l},
    \label{eq:h_l}
\end{eqnarray}
where we have used $\mid\vec{d} - \vec{r}_{n}\mid \approx d$ for amplitudes and $\mid\vec{d} - \vec{r}_{n}\mid \approx d - \mid\vec{d}\cdot\vec{r}_{n}\mid$ for phases~\cite{OAM_study}.

When $N\rightarrow\infty$, we have
\begin{eqnarray}
   &&{\hspace{-0.6cm}}\sum_{n=1}^{N}e^{j\frac{2 \pi \mid\vec{d}\cdot\vec{r}_{n}\mid }{\lambda}}e^{j \frac{2 \pi (n-1)}{N} l}\nonumber\\
   &=&\sum_{n=1}^{N}e^{j\frac{2 \pi}{\lambda} R \sin{\theta} \cos{\varphi}} e^{j \frac{2 \pi (n-1)}{N} l}\nonumber\\
  & \approx & \! \frac{N e^{j \theta l}}{2\pi} \int_{0}^{2\pi} e^{j\frac{2 \pi}{\lambda} R \sin{\theta} \cos{\varphi}^{'}} e^{-j \varphi^{'} l}d \varphi^{'}\nonumber\\
   &= & Nj^{-l} e^{j \varphi l} J_{l}\left(\frac{2\pi}{\lambda}R\sin{\theta}\right),
\end{eqnarray}
where $R$ denotes the radius of the UCA antenna, $\theta$ denotes the included angle between the normal line of the transmit UCA and the line from the center of the OAM-receiver to the center of OAM-transmitter, and 
\begin{eqnarray}
    J_{l}(x)=\frac{1}{2\pi j^{-l}}\int_{0}^{2\pi} e^{j(x\cos(\varphi^{\prime})-l\varphi^{\prime})}d\varphi^{\prime},
\end{eqnarray}
is the first kind Bessel function with order $l$.
Thus, the expression of channel amplitude gain $h_{l}$ can be re-written as follows:
\begin{eqnarray}
    h_{l}=\frac{\beta \lambda N j^{-l}}{4 \pi d} e^{-j \frac{2\pi}{\lambda} d} e^{j\varphi l} J_{l}\left(\frac{2\pi}{\lambda}R\sin{\theta}\right).\label{eq:h_l}
\end{eqnarray}
Based on Eq.~\eqref{eq:h_l}, we can find that $h_{l}$ increases as the number of array-elements increases.

\subsection{Received Signal}

Replacing $l$ by $l_{u}$ and $l_{u,k}$, respectively, in Eq.~\eqref{eq:h_l}, and substituting $h_{l_{u}}$ and $h_{l_{u,k}}$ into Eq.~\eqref{eq:r_MH} as well as multiplying $r_{1}(u, t)$ with $e^{j\varphi l_{u}}$, we can obtain the de-hopping signal, denoted by $\tilde{r}_{1}(u,t)$, for the $u$-th hop as follows:
\begin{eqnarray}
    \tilde{r}_{1}(u,t)=r_{1}(u,t)e^{j\varphi l_{u}}.
\end{eqnarray}

Our goal aims to recover the transmit signal for the desired user. However, $\tilde{r}_{1}(u,t)$ carries OAM-mode and thus needs to be decomposed. Using the integrator, we can obtain the decomposed signal, denoted by $r_{1}^{\prime}(u,t)$, specified in Eq.~\eqref{eq:M_integrator}, where $(\cdot)^{*}$ represents the complex conjugate operation, $\tilde{n}(u,t)$ represents the received noise for the $u$-th hop after OAM-mode decomposition, and $D_{u}\subseteq \{1, 2, \cdots, K\}$.
\setcounter{TempEqCnt}{\value{equation}} 
\setcounter{equation}{15} 
\begin{figure*}[ht]
\begin{eqnarray}
P_{e}(V, D_{v}|U)\hspace{-0.3cm} &=& \hspace{-0.3cm} \left\{
\begin{array}{lll}
m^{mU}\Bigg\{\sum\limits_{v_{1}=0}^{U-V-1}\sum\limits_{u_{1}=1}^{m\left(U-V\right)
}\frac{2^{1-2\left(U-V\right)}P_{u_{1}}c_{v_{1}}\Gamma\left[m\left(U-V\right)-u_{1}+v_{1}+2\right]}{\zeta^{m\left(U-
V\right)-u_{1}+1}\Gamma\left[m\left(U-V\right)-u_{1}+1\right]\left(\mu +\frac{m}{\zeta}\right)^{m\left(U-V
\right)-u_{1}+v_{1}+2}}\\
+\sum\limits_{v_{1}=0}^{a-1}\sum\limits_{u_{2}=1}^{ma}\frac{2^{1-2a}Q_{u_{2}}c_{v_{1}}\Gamma\left(ma-u_{2}+v_{1}+2\right)}{\overline{\delta}_{L}
^{ma-u_{2}+1}\Gamma\left(ma-u_{2}+1\right)\left(\mu+\frac{m}{\overline{\delta}_{L}}\right)^{ma-u_{2}+v_{1}+2}}\\
+\sum\limits_{v_{1}=0}^{V-a-1}\sum\limits_{v=1}^{V-a}\sum\limits_{u_{3}=1}^{m}\frac{2^{1-2\left(V-a\right)}W_{vu_{3}}
c_{v_{1}}\Gamma\left(m-u_{3}+v_{1}+2\right)}{\overline{\delta}_{v}
^{m-u_{3}+1}\Gamma\left(m-u_{3}+1\right)\left(\mu+\frac{m}{\overline{\delta}_{v}}\right)^{m-u_{3}+v_{1}+2}}
\Bigg\}, \hspace{4.5cm} a \geq 1;
\vspace{0.2cm}\\
m^{mU}\Bigg\{\sum\limits_{v_{1}=0}^{U-V-1}\sum\limits_{u_{1}=1}^{m\left(U-V\right)
}\frac{2^{1-2\left(U-V\right)}P_{u_{1}}c_{v_{1}}\Gamma\left[m\left(U-V\right)-u_{1}+v_{1}+2\right]}{\zeta^{m\left(U-
V\right)-u_{1}+1}\Gamma\left[m\left(U-V\right)-u_{1}+1\right]\left(\mu +\frac{m}{\zeta}\right)^{m\left(U-V
\right)-u_{1}+v_{1}+2}}\\
+\sum\limits_{v_{1}=0}^{V-1}\sum\limits_{v=1}^{V}\sum\limits_{u_{3}=1}^{m}\frac{2^{1-2V}W_{vu_{3}}
c_{v_{1}}\Gamma\left(m-u_{3}+v_{1}+2\right)}{\overline{\delta}_{v}
^{m-u_{3}+1}\Gamma\left(m-u_{3}+1\right)\left(\mu+\frac{m}{\overline{\delta}_{v}}\right)^{m-u_{3}+v_{1}+2}}
\Bigg\},\hspace{5cm} a = 0.
\end{array}
\right.
\label{eq:BER_u_MH_all}
\end{eqnarray}
\hrulefill
\vspace{-10pt}
\end{figure*}
\setcounter{equation}{\value{TempEqCnt}}

Based on Eq.~\eqref{eq:M_integrator}, we can calculate the signal-to-noise ratios (SNRs) for the scenarios with $l_{u,k} \neq l_{u}$ and signal-to-interference-plus-noise ratios (SINRs) for the scenarios with $l_{u,k} = l_{u}$. We denote by $E_{h}$ the transmit power for each hop. For the scenario with $l_{u,k} \neq l_{u}$, the received instantaneous SNR, denoted by $\gamma_{u}$, after OAM-mode decomposition for the $u$-th hop can be expressed as follows:
\setcounter{equation}{10}
\begin{eqnarray}
   \gamma_{u}=\frac{E_{h}h_{l_{u}}^{2}}{\sigma_{l_{u}}^{2}},
\end{eqnarray}
where $\sigma_{l_{u}}^{2}$ is the variance of the received noise after OAM-mode decomposition for the $u$-th hop corresponding to the $\l_{u}$-th OAM-mode. Under the scenario with $l_{u} \neq l_{u,k}$, we assume the average SNRs for each hop  are the same. Thus, the average SNR, denoted by $\zeta$, can be expressed as follows:
\begin{eqnarray}
    \zeta=E_{h}\mathbb{E}\left(\frac{h_{l_{u}}^{2}}{\sigma_{l_{u}}^{2}}\right),
\end{eqnarray}
where $\mathbb{E}(\cdot)$ represents the expectation operation.

For the scenario with $l_{u,k} = l_{u}$, we assume that there are $V$ ($1 \leq V \leq U$) hops for the signal of desired user jammed by interfering users. We also assume that there are corresponding $D_{v} \ (1\leq D_{v}\leq K, 1\leq v \leq V)$ interfering users for the $v$-th hop. The received instantaneous SINR, denoted by $\delta_{v}$, with $D_{v}$ interfering users after OAM-mode decomposition for the $v$-th hop in MH communications can be derived as follows:

\begin{eqnarray}
  \delta_{v}=\frac{h_{\tilde{l}_{v}}^{2}E_{h}}{\sigma_{\tilde{l}_{v}}^{2}+\sum\limits_{k=1}^{D_{v}}E_{h} h_{\tilde{l}_{v},k}^{2}},
 \end{eqnarray}
where $h_{\tilde{l}_{v}}$ is the channel gain corresponding to the channel from the desired OAM-transmitter to the OAM-receiver for the $v$-th hop and can be obtained by replacing $l$ by $\tilde{l}_{v}$ in Eq.~\eqref{eq:h_l}, $h_{\tilde{l}_{v,k}}$ is the channel gain corresponding to the channel from the interfering users' OAM-transmitter to the OAM-receiver for the $v$-th hop and can be obtained  by replacing $l$ by $\tilde{l}_{v,k}$ in Eq.~\eqref{eq:h_l}, and $\sigma_{\tilde{l}_{v}}^{2}$ is the variance of the received noise after OAM-mode decomposition for the $v$-th hop corresponding to the $\tilde{l}_{v}$-th OAM-mode.

For the scenario with $l_{u,k} = l_{u}$, the average SINR, denoted by $\overline{\delta}_{v}$, with $D_{v}$ interfering users corresponding to the $v$-th hop can be expressed as follows:
 \begin{eqnarray}
   \overline{\delta}_{v}=\mathbb{E}\left(\frac{E_{h} h_{\tilde{l}_{v}}^{2} }{\sum\limits_{k=1}^{D_{v}} E_{h} h_{\tilde{l}_{v},k}^{2}+\sigma_{\tilde{l}_{v}}^{2}}\right).
 \end{eqnarray}

We employ the equal gain combining (EGC) diversity reception. Thus, the received instantaneous SINR, denoted by $\gamma_{s}$, for the desired user for $U$ hops at the output of EGC diversity reception can be obtained as follows:
\begin{eqnarray}
   \gamma_{s}=\frac{E_{h}\left(\sum\limits_{u=1}^{U-V}h_{l_{u}}^{2}+\sum\limits_{v=1}^{V}h_{\tilde{l}_{v}}^{2}\right)}
   {\sum\limits_{u=1}^{U-V}\sigma_{l_{u}}^{2}+\sum\limits_{v=1}^{V}
   \left(\sum\limits_{k=1}^{D_{v}}E_{h}h_{\tilde{l}_{v,k}}^{2}+\sigma_{\tilde{l}_{v,k}}^{2}\right)}.
\end{eqnarray}

In the MH system, the complexity of MH receiver mainly depends on OAM-receiver, mode synthesizer, integrator, and EGC. The complexity of FH receiver mainly depends on receive antenna, frequency synthesizer, and EGC. The complexity of mode synthesizer in the MH system is similar to the complexity of frequency synthesizer in the FH system. Also, the complexity of EGC used in the MH system is similar to the complexity of EGC used in the FH system. Although the FH receiver uses single receive antenna, the MH system uses OAM-receiver based UCA which can be considered as single radio frequency chain antenna. In addition, the MH system adds a simple integrator.

\subsection{Performance Analysis}

To analyze the performance of our developed MH scheme, we employ binary differential phase shift keying (DPSK) and binary non-coherent frequency shift keying (FSK) modulation. We introduce a constant denoted by $\mu$. If $\mu=1$, it means that we employ binary DPSK modulation. If $\mu=1/2$, it implies that we employ binary non-coherent FSK modulation. In addition, Nakagami-$m$ fading can be used to analyze the system performance in radio vortex wireless communications~\cite{2013_fading}.

Then, we assume that the number of interfering users is $L$ ($ 1\leq L \leq K$) for $a$ $(0\leq a \leq V)$ hops while the numbers of interfering users for the other $(V-a)$ hops are different. We denote by $\overline{\delta}_{L}$ the average SINR with $L$ interfering users.

Then, we have the following Theorem 1.

\emph{Theorem 1}: The average BER, denoted by $P_{e}(V, D_{v}|U)$, of MH scheme given $V$ hops jammed by corresponding $D_{v}$ interfering users with Nakagami-$m$ fading channel is given by Eq.~\eqref{eq:BER_u_MH_all}.

\begin{proof}
See Appendix~\ref{Appendix:_A}.
\end{proof}
When $V=U$, $a=0$, and $m=1$, Eq.~\eqref{eq:phi_MU} can be reduced to
\setcounter{equation}{16}
\begin{eqnarray}
     \Phi_{\gamma_{s}}(w)=\sum_{v=1}^{U}M_{u} y_{v},
\end{eqnarray}
where
\begin{subnumcases}
{} M_{v}=\prod_{\substack{i=1 \\ i\neq v}}^{U}\frac{\overline{\delta}_{v}}{\overline{\delta}_{v}-\overline{\delta}_{i}};
\\
y_{v}=\frac{1}{1-j w \overline{\delta}_{v}}.
\end{subnumcases}
Thus, when $V=U$, $a=0$, and $m=1$, the PDF of $\gamma_{s}$ can be reduced to
\begin{eqnarray}
    p_{\gamma_{s}}(\gamma_{s})=\sum_{v=1}^{U} M_{v} \frac{e^{-\gamma_{s}/\overline{\delta}_{v}}}{ \overline{\delta}_{v}}.\label{eq:P_gamma_s_MH_spec}
\end{eqnarray}
Then, when $V=U$, $a=0$, and $m=1$, the BER $P_{e}(V, D_{v}|U)$ can be reduced to
\begin{eqnarray}
 &&\hspace{-0.5cm}P_{e}(V, D_{v}|U)\nonumber\\
 &&\hspace{-0.5cm}=\left(\frac{1}{2}\right)^{2U-1} \sum_{v=1}^{U} \sum_{v_{1}=0}^{U-1}
 \frac{M_{v}\Gamma(v_{1}-1) c_{v_{1}}}{\overline{\delta}_{v}}\left(\frac{\overline{\delta}_{v}}{1+\mu\overline{\delta}_{v}}\right)^{v_{1}+1}.\nonumber\\
 \label{eq:BER_u_MU_k}
\end{eqnarray}

For the scenario with $l_{u} \neq l_{u,k}$ corresponding to $U$ hops, we assume that interfering users use different OAM-modes from the desired user. 

Then, we have the following Theorem 2.

\emph{Theorem 2}: The average BER, denoted by $P_{e}(U)$, of MH scheme for the scenario with $l_{u} \neq l_{u,k}$ corresponding to $U$ hops is given by

\begin{eqnarray}
  P_{e}(U)\hspace{-0.2cm}&=&\hspace{-0.2cm}\frac{2^{1-2U}}{\Gamma(mU)}\left(\frac{m}{m+\mu\zeta}\right)^{mU}\nonumber\\
  &&\sum\limits_{v_{1}=0}^{U-1}c_{v_{1}}\Gamma(mU+v_{1})\left(\frac{\zeta}{m+\mu\zeta}\right)^{v_{1}}.
  \label{eq:BER_P_e_U}
\end{eqnarray}
\begin{proof}
  See Appendix~\ref{Appendix:_C}.
\end{proof}

When the OAM-mode of interfering users is different from that of the desired user in MH communications, the system is equivalent to the single user system. Thus, the average BER of single user system can be expressed as Eq.~\eqref{eq:BER_P_e_U}.
\setcounter{TempEqCnt}{\value{equation}} 
\setcounter{equation}{27} 
\begin{figure*}[ht]
\begin{subnumcases}
{r_{2}^{\prime}(u,t)=\frac{1}{2\pi} \int_{0}^{2 \pi}\tilde{r}_{2}(u,t) \left(e^{j 2\varphi l_{u}}\right)^{*} d \varphi =}
h_{l_{u}}s(t)\varepsilon(t-u t_{h})\cos^{2}(2 \pi f_{u}t+ \alpha_{u})+\tilde{n}(u,t),\ \ \ \ \ \  l_{u,k}\neq l_{u};\label{eq:r_mf_m_MU_a}\\
\sum_{k=1}^{ D_{u}}h_{l_{u},k}s_{k}(t)\varepsilon(t\!\!-\!\!u t_{h})\cos(2 \pi f_{u,k}t\!+\! \alpha_{u,k})\cos(2 \pi f_{u}t\!+\! \alpha_{u})\nonumber\\
+h_{l_{u}}s(t)\varepsilon(t-u t_{h})\cos^{2}(2 \pi f_{u}t+\alpha_{u})+ \tilde{n}(u,t),\ \ \ \ l_{u,k}= l_{u}.
\label{eq:r_mf_m_MU_b}
\end{subnumcases}
\hrulefill
\end{figure*}
\setcounter{equation}{\value{TempEqCnt}}
\setcounter{TempEqCnt}{\value{equation}} 
\setcounter{equation}{28} 
\begin{figure*}[ht]
\begin{subnumcases}
 {y(u,t)=}
h_{l_{u}}s(t)\varepsilon(t-u t_{h})+\tilde{n}(u,t),\ \hspace{4cm} \ f_{u,k}\neq f_{u}, l_{u,k}\neq l_{u};\label{eq:r_mf_f_MU_a}\\
h_{l_{u}}s(t)\varepsilon(t-u t_{h})+\tilde{n}(u,t),\ \hspace{4cm} \ f_{u,k}= f_{u}, l_{u,k} \neq l_{u};\label{eq:r_mf_f_MU_b}\\
h_{l_{u}}s(t)\varepsilon(t-u t_{h})+\tilde{n}(u,t),\ \hspace{4cm} \ f_{u,k}\neq  f_{u}, l_{u,k}= l_{u};
\label{eq:r_mf_f_MU_c}\\
h_{l_{u}}s(t)\varepsilon(t-u t_{h})+ \tilde{n}(u,t)+\sum_{k=1}^{D_{u}}h_{l_{u},k}s_{k}(t)\varepsilon(t-u t_{h}),\ \  f_{u,k}= f_{u}, l_{u,k}= l_{u}.
\label{eq:r_mf_f_MU_d}
\end{subnumcases}
\hrulefill
\end{figure*}
\setcounter{equation}{\value{TempEqCnt}}

Then, we denote by $P_{0}$ the probability that the signal of desirable user is jammed by an interfering user. Assuming that the probability of an OAM-mode carried by transmit signal for each hop is equal to $1/N$ in MH communications, the probability $P_{0}$ is equal to $1/N$ for each hop. Thereby, the probability, denoted by $p(U)$, for the scenario with $l_{u} \neq l_{u,k}$ corresponding to $U$ hops can be obtained as follows:
\begin{eqnarray}
    p(U)=(1-P_{0})^{KU}.\label{eq:BER_no_MAI}
\end{eqnarray}

Also, we can calculate the probability, denoted by $p(V|U)$, given $V$ hops jammed by interfering users for $U$ hops as follows:
\begin{eqnarray}
p(V|U)\hspace{-0.2cm}&=&\hspace{-0.2cm}\underbrace{\sum_{D_{1}=1}^{K}\sum_{D_{2}=1}^{K} \cdots \sum_{D_{V}=1}^{K}}_{V-fold}\dbinom{U}{V}
    (1-P_{0})^{K(U-V)}\nonumber\\
    &&\hspace{-0.2cm}\hspace{0.2cm}\prod_{v=1}^{V}\dbinom{K}{D_{v}} P_{0}^{D_{v}} \left(1-P_{0}\right)^{K-D_{v}}.\label{eq:BER_u_MU}
\end{eqnarray}

Then, we can calculate the average BER, denoted by $P_{s}$, considering all possible cases in MH communications as follows:
\begin{eqnarray}
P_{s}=p(U)\ P_{e}(U)+\sum_{V=1}^{U}p(V|U)P_{e}(V,  D_{v}|U).\label{eq:BER_MU}
\end{eqnarray}
Substituting Eqs.~\eqref{eq:BER_u_MH_all},~\eqref{eq:BER_P_e_U},~\eqref{eq:BER_no_MAI}, and~\eqref{eq:BER_u_MU} into Eq.~\eqref{eq:BER_MU}, we can obtain the average BER for all possible cases in MH communications.

Observing the BER of MH scheme, we can find that the average SINR, the number of interfering users, the number of OAM-modes, and the number of hops make impact on the value of BER. Although the BER has complex form, we can directly obtain some results. First, the BER monotonically increases as the number of interfering users increases.
Second, the BER monotonically decreases as the average SINR increases. Thus, increasing the transmit power and mitigating interference can decrease the BER.
In addition, the BER monotonically decreases as the number of OAM-modes increases. Generally, the number of OAM-modes mainly depends on the number of array-elements used by the OAM-transmitter. Hence, increasing the number of array-elements corresponding to the OAM-transmitter can decrease the BER in the MH system.
Also, the BER monotonically decreases as the number of hops increases. Thus, increasing the number of hops can decrease the BER in the MH system.

\section{MFH Scheme}\label{sec:MFH}
For our developed MFH communications, the transmit signal, denoted by $x_{2}(u,t)$, for the desired user corresponding to the $u$-th hop can be expressed as follows:
\begin{align}
  x_{2}(u,t)=s(t)\varepsilon(t-ut_{h})e^{j \varphi l_{u}}\cos(2\pi f_{u}t+\alpha_{u}),
\end{align}
where $\alpha_{u}$ $(0 \leq \alpha_{u} \leq 2\pi)$ denotes the initial phase corresponding to the $u$-th hop. We denote by $f_{u,k}$ the carrier frequency and $\alpha_{u,k}$ $(0 \leq \alpha_{u,k} \leq 2\pi)$ the initial phase distributed of the $k$-th interfering user for the $u$-th hop, respectively. Thus, the received signal, denoted by $r_{2}(u,t)$, for the desired user corresponding to the $u$-th hop can be derived as follows:
\begin{eqnarray}
  r_{2}(u,t)\!\!&\hspace{-0.3cm}=\hspace{-0.3cm}& h_{l_{u}}x_{2}(u,t)+n(u,t)\nonumber\\
  &&\!\!+\!\!\sum_{k=1}^{K}h_{l_{u},k}s_{k}(t)\varepsilon(t\!\!-\!\!ut_{h})e^{j \varphi l_{u,k}}\cos(2 \pi f_{u,k}t \!\!+ \!\!\alpha_{u,k}).\nonumber\\
 \end{eqnarray}
Then, multiplying $r_{2}(u,t)$ with $e^{j \varphi l_{u}}$ and $\cos(2\pi f_{u}t+\alpha_{u})$, we can obtain the de-hopping signal, denoted by $\tilde{r}_{2}(u,t)$, for the $u$-th hop as follows:
\begin{eqnarray}
   \tilde{r}_{2}(u,t)=r_{2}(u,t)e^{j \varphi l_{u}}\cos(2\pi f_{u}t+\alpha_{u}).
\end{eqnarray}
Using the integrator, we can obtain the decomposed signal, denoted by $r_{2}^{\prime}(u,t)$, specified in Eqs.~\eqref{eq:r_mf_m_MU_a} and ~\eqref{eq:r_mf_m_MU_b}. Then, using the low pass filter after OAM-mode decomposition, we can obtain the received signal, denoted by $y(u,t)$, for the desired user corresponding to the $u$-th hop as Eqs.~\eqref{eq:r_mf_f_MU_a},~\eqref{eq:r_mf_f_MU_b},~\eqref{eq:r_mf_f_MU_c}, and~\eqref{eq:r_mf_f_MU_d}. Clearly, there are four cases, which are described in the following.

Case 1: If both OAM-modes and carrier frequencies of $K$ interfering users are different from that of desired user for each mode/frequency hop, the received signal can be obtained as the right hand of the Eqs.~\eqref{eq:r_mf_m_MU_a} and ~\eqref{eq:r_mf_f_MU_a}. Thus, we can obtain that the interfering signals can be entirely removed.

Case 2: If carrier frequencies of the $K$ interfering users are the same with that of desired user while OAM-modes of the $K$ interfering users are different from that of the desired user for each hop, we can obtain the signal as the right hand of the Eqs.~\eqref{eq:r_mf_m_MU_a} and ~\eqref{eq:r_mf_f_MU_b}. Clearly, the interfering signals can also be entirely removed.

Case 3: If the OAM-modes of $K$ interfering users are the same modes with that of desired user while carrier frequencies of interfering users are different from that of the desired user for each hop, the received signal can be expressed as the right hand of Eqs.~\eqref{eq:r_mf_m_MU_b} and ~\eqref{eq:r_mf_f_MU_c}. In this case, the integrator mentioned above doesn't work for interfering signals while the low pass filter can filter out interfering signals.

Case 4: If both the carrier frequencies and OAM-modes of $K$ interfering users are the same with that of the desired user, the received signal can be given as the right hand of the Eqs.~\eqref{eq:r_mf_m_MU_b} and ~\eqref{eq:r_mf_f_MU_d}. The integrator and low pass filter cannot cancel interfering signals.

Observing the above four cases, we can know that the interfering signals make impact on the performance of the MFH system only when both the OAM-modes and carrier frequencies of interfering users are the same with that of the desired user.
The interfering signals can be canceled by the integrator or low pass filter for the first three cases. Only Case 4 degrades the performance of system.

For Case 1, 2 and 3, the instantaneous SNR, denoted by $\rho_{u}$, after OAM decomposition for the $u$-th hop in MFH communications, can be expressed as follows:
\setcounter{equation}{29}
\begin{eqnarray}
    \rho_{u}=\frac{h_{l_{u}}^{2}E_{h}}{\Omega_{l_{u}}^{2}},\label{eq:rho}
\end{eqnarray}
where $\Omega_{l_{u}}^{2}$ denotes the variance of received noise after OAM decomposition for the $u$-th hop in MFH communications.

For Case 4, the instantaneous SINR, denoted by $\varrho_{v}$, with $D_{v}$ interfering users after OAM decomposition for the $v$-th hop in MFH communications, can be expressed as follows:
\begin{eqnarray}
    \varrho_{v}=\frac{h_{\tilde{l}_{v}}^{2}E_{h}}{\Omega_{\tilde{l}_{v}}^{2}+\sum\limits_{u=1}^{D_{v}}E_{h}h_{\tilde{l}_{v,k}}^{2}},\label{eq:varrho}
\end{eqnarray}
where $\Omega_{\tilde{l}_{v}}^{2}$ is the variance of received noise after OAM decomposition corresponding to the $v$-th hop for Case 4 in MFH communications.

In MFH communications, signal can be transmitted in both angle/mode domain and frequency domain simultaneously by jointly using MH and FH schemes. Thus, MH and FH are mutually independent to each other. Hence, the processing gain~\cite{1982_processing_gain} of MFH scheme is the product of the processing gains of FH scheme and MH scheme. We denote by $G$ (which approximates to $Q$) the processing gain of FH scheme. Given a fix transmit SNR, the receive SNR of MFH scheme is $G$ times that of MH scheme. Thus, we can re-write Eq.~\eqref{eq:rho} as follows:
\begin{eqnarray}
    \rho_{u}=G \gamma_{u}.\label{eq:rho_G}
\end{eqnarray}
Also, we can re-write Eq.~\eqref{eq:varrho} as follows:
\begin{eqnarray}
    \varrho_{v}=\delta_{v}+\frac{(G-1)\Omega_{\tilde{l}_{v}}^{2}}{\Omega_{\tilde{l}_{v}}^{2}+\sum\limits_{u=1}^{D_{v}}E_{h}h_{\tilde{l}_{v,k}}^{2}}\delta_{v}.
    \label{eq:varrho_G}
\end{eqnarray}
Based on Eq.~\eqref{eq:varrho_G}, we can find that the received average SNR or SINR of MFH scheme is always larger than that of MH scheme. 
For Case 1, 2, and 3, the average SNR, denoted by $\xi$, can be calculated as follows:
\begin{eqnarray}
    \xi=G \zeta.
\end{eqnarray}
For Case 4, the average SINR, denoted by $\overline{\varrho}_{v}$, with $D_{v}$ interfering users for the $v$-th hop can be derived as follows:
\begin{eqnarray}
    \overline{\varrho}_{v}=\overline{\delta}_{v}+\mathbb{E}\left[\frac{(G-1)\Omega_{\tilde{l}_{v}}^{2}}{\Omega_{\tilde{l}_{v}}^{2}+\sum\limits_{u=1}^{D_{v}}
    E_{h} h_{\tilde{l}_{v},k}^{2}}\right]\overline{\delta}_{v}.\label{eq:average_varrho}
\end{eqnarray}
Replacing $D_{v}$ by $L$ in Eq.~\eqref{eq:average_varrho}, we can obtain the average SINR, denoted by $\overline{\varrho}_{L}$, with $L$ interfering users.

\setcounter{equation}{\value{TempEqCnt}}
\setcounter{TempEqCnt}{\value{equation}} 
\setcounter{equation}{37} 
\begin{figure*}[ht]
\begin{eqnarray}
P_{e}^{\prime}(V, D_{v}|U)\hspace{-0.3cm} &=& \hspace{-0.3cm} \left\{
\begin{array}{lll}
m^{mU}\Bigg\{\sum\limits_{v_{1}=0}^{U-V-1}\sum\limits_{u_{1}=1}^{m\left(U-V\right)
}\frac{2^{1-2\left(U-V\right)}P_{u_{1}}c_{v_{1}}\Gamma\left[m\left(U-V\right)-u_{1}+v_{1}+2\right]}{\xi^{m\left(U-
V\right)-u_{1}+1}\Gamma\left[m\left(U-V\right)-u_{1}+1\right]\left(\mu+\frac{m}{\xi}\right)^{m\left(U-V
\right)-u_{1}+v_{1}+2}}\\
+\sum\limits_{v_{1}=0}^{a-1}\sum\limits_{u_{2}=1}^{ma}\frac{2^{1-2a}Q_{u_{2}}c_{v_{1}}\Gamma\left(ma-u_{2}+v_{1}+2\right)}{\overline{\varrho}_{L}
^{ma-u_{2}+1}\Gamma\left(ma-u_{2}+1\right)\left(\mu+\frac{m}{\overline{\varrho}_{L}}\right)^{ma-u_{2}+v_{1}+2}}\\
+\sum\limits_{v_{1}=0}^{V-a-1}\sum\limits_{v=1}^{V-a}\sum\limits_{u_{3}=1}^{m}\frac{2^{1-2\left(V-a\right)}W_{vu_{3}}
c_{v_{1}}\Gamma\left(m-u_{3}+v_{1}+2\right)}{\overline{\varrho}_{v}
^{m-u_{3}+1}\Gamma\left(m-u_{3}+1\right)\left(\mu+\frac{m}{\overline{\varrho}_{v}}\right)^{m-u_{3}+v_{1}+2}}
\Bigg\}, \hspace{4.5cm}a \geq 1;
\vspace{0.2cm}\\
m^{mU}\Bigg\{\sum\limits_{v_{1}=0}^{U-V-1}\sum\limits_{u_{1}=1}^{m\left(U-V\right)
}\frac{2^{1-2\left(U-V\right)}P_{u_{1}}c_{v_{1}}\Gamma\left[m\left(U-V\right)-u_{1}+v_{1}+2\right]}{\xi^{m\left(U-
V\right)-u_{1}+1}\Gamma\left[m\left(U-V\right)-u_{1}+1\right]\left(\mu+\frac{m}{\xi}\right)^{m\left(U-V
\right)-u_{1}+v_{1}+2}}\\
+\sum\limits_{v_{1}=0}^{V-1}\sum\limits_{v=1}^{V}\sum\limits_{u_{3}=1}^{m}\frac{2^{1-2V-}W_{vu_{3}}
c_{v_{1}}\Gamma\left(m-u_{3}+v_{1}+2\right)}{\overline{\varrho}_{v}
^{m-u_{3}+1}\Gamma\left(m-u_{3}+1\right)\left(\mu+\frac{m}{\overline{\varrho}_{v}}\right)^{m-u_{3}+v_{1}+2}}
\Bigg\}, \hspace{5cm}a = 0.
\end{array}
\right.
 \label{eq:BER_u_MF_all}
\end{eqnarray}
\hrulefill
\end{figure*}\setcounter{equation}{\value{TempEqCnt}}
\setcounter{TempEqCnt}{\value{equation}} 
\setcounter{equation}{38} 
\begin{figure*}[ht]
\begin{eqnarray}
P_{s}^{\prime}=\left\{
\begin{array}{lll}
\hspace{-0.2cm}\left(1-P_{1}\right)^{KU}\frac{2^{1-2U}}{\Gamma(mU)}\left(\frac{m}{m+G\mu\zeta}\right)^{mU}
  \sum\limits_{v_{1}=0}^{U-1}c_{v_{1}}\Gamma(mU+v_{1})\left(\frac{G\zeta}{m+G\mu\zeta}\right)^{v_{1}}
  \\
\hspace{-0.2cm}+\left[\sum\limits_{D_{1}=1}^{K}\sum\limits_{D_{2}=1}^{K} \cdots \sum\limits_{D_{V}=1}^{K}\dbinom{U}{V}(1-P_{1})^{K(U-V)}\prod\limits_{v=1}^{V}\dbinom{K}{D_{v}} P_{1}^{D_{i}} \left(1-P_{1}\right)^{K-D_{v}}\right]
  \\
\hspace{-0.2cm}m^{mU}\Bigg\{\sum\limits_{v_{1}=0}^{U-V-1}\sum\limits_{u_{1}=1}^{m\left(U-V\right)
}\frac{2^{1-2\left(U-V\right)}P_{u_{1}}c_{v_{1}}\Gamma\left[m\left(U-V\right)-u_{1}+v_{1}+2\right]}{(G\zeta)^{m\left(U-
V\right)-u_{1}+1}\Gamma\left[m\left(U-V\right)-u_{1}+1\right]\left(\mu+\frac{m}{G\zeta}\right)^{m\left(U-V
\right)-u_{1}+v_{1}+2}}\\
\hspace{-0.2cm}+\sum\limits_{v_{1}=0}^{a-1}\sum\limits_{u_{2}=1}^{ma}\frac{2^{1-2a}Q_{u_{2}}c_{v_{1}}\Gamma\left(ma-u_{2}+v_{1}+2\right)}{\overline{\varrho}_{L}
^{ma-u_{2}+1}\Gamma\left(ma-u_{2}+1\right)\left(\mu+\frac{m}{\overline{\varrho}_{L}}\right)^{ma-u_{2}+v_{1}+2}}
+\sum\limits_{v_{1}=0}^{V-a-1}\sum\limits_{v=1}^{V-a}\sum\limits_{u_{3}=1}^{m}\frac{2^{1-2\left(V-a\right)}W_{vu_{3}}
c_{v_{1}}\Gamma\left(m-u_{3}+v_{1}+2\right)}{\overline{\varrho}_{v}
^{m-u_{3}+1}\Gamma\left(m-u_{3}+1\right)\left(\mu+\frac{m}{\overline{\varrho}_{v}}\right)^{m-u_{3}+v_{1}+2}}
\Bigg\}, \\
\hspace{13.8cm}a \geq 1;
\\
\hspace{-0.2cm}\left(1-P_{1}\right)^{KU}\frac{2^{1-2U}}{\Gamma(mU)}\left(\frac{m}{m+G\mu\zeta}\right)^{mU}
  \sum\limits_{v_{1}=0}^{U-1}c_{v_{1}}\Gamma(mU+v_{1})\left(\frac{G\zeta}{m+G\mu\zeta}\right)^{v_{1}}
  \\
\hspace{-0.2cm}  +\left[\sum\limits_{D_{1}=1}^{K}\sum\limits_{D_{2}=1}^{K} \cdots \sum\limits_{D_{V}=1}^{K}\dbinom{U}{V}(1-P_{1})^{K(U-V)}\prod\limits_{v=1}^{V}\dbinom{K}{D_{v}} P_{1}^{D_{i}} \left(1-P_{1}\right)^{K-D_{v}}\right]
  \\
\hspace{-0.2cm}m^{mU}\Bigg\{\sum\limits_{v_{1}=0}^{U-V-1}\sum\limits_{u_{1}=1}^{m\left(U-V\right)
}\frac{2^{1-2\left(U-V\right)}P_{u_{1}}c_{v_{1}}\Gamma\left[m\left(U-V\right)-u_{1}+v_{1}+2\right]}{(G\zeta)^{m\left(U-
V\right)-u_{1}+1}\Gamma\left[m\left(U-V\right)-u_{1}+1\right]\left(\mu+\frac{m}{G\zeta}\right)^{m\left(U-V
\right)-u_{1}+v_{1}+2}}\\
\hspace{-0.2cm}+\sum\limits_{v_{1}=0}^{V-1}\sum\limits_{v=1}^{V}\sum\limits_{u_{3}=1}^{m}\frac{2^{1-2V}W_{vu_{3}}
c_{v_{1}}\Gamma\left(m-u_{3}+v_{1}+2\right)}{\overline{\varrho}_{v}
^{m-u_{3}+1}\Gamma\left(m-u_{3}+1\right)\left(\mu+\frac{m}{\overline{\varrho}_{v}}\right)^{m-u_{3}+v_{1}+2}}
\Bigg\}, \hspace{6cm}a =0.
\end{array}
\right.
 \label{eq:BER_P_all_MFH}
\end{eqnarray}
\hrulefill
\end{figure*}
\setcounter{equation}{\value{TempEqCnt}}
\begin{eqnarray}
\setcounter{equation}{36}
  P_{e}^{\prime}(U)\hspace{-0.2cm}&=&\hspace{-0.2cm}\frac{2^{1-2U}}{\Gamma(mU)}\left(\frac{m}{m+G\mu\zeta}\right)^{mU}\nonumber\\
  &&\sum\limits_{v_{1}=0}^{U-1}c_{v_{1}}\Gamma(mU+v_{1})\left(\frac{G\zeta}{m+G\mu\zeta}\right)^{v_{1}}.\label{eq:BER_mf}
\end{eqnarray}
We denote by $\rho_{s}$ the received instantaneous SINR for $U$ hops at the output of EGC diversity reception in MFH communications. Then, we have
\begin{eqnarray}
   \rho_{s}=\frac{E_{h}\left(\sum\limits_{u=1}^{U-V}h_{l_{u}}^{2}+\sum\limits_{v=1}^{V}h_{\tilde{l}_{v}}^{2}\right)}
   {\sum\limits_{u=1}^{U-V}\Omega_{l_{u}}^{2}+\sum\limits_{v=1}^{V}\left(\sum\limits_{k=1}^{D_{v}}E_{h}h_{\tilde{l}_{v,k}}+\Omega_{\tilde{l}_{v,k}}^{2}\right)}.
\end{eqnarray}


Similar to the analyses of the BER in MH communications, the BER, denoted by $P_{e}^{\prime}(U)$, for $U$ hops without interfering users in MFH communications can be derived as follows:

We denote by $P_{e}^{\prime}(V, D_{v}|U)$ the average BER of MFH system given $V$ hops jammed by corresponding $D_{v}$ interfering users with Nakagami-$m$ fading channel. Then, we can obtain the average BER $P_{e}^{\prime}(V,  D_{v}|U)$ in MFH communications as Eq.~\eqref{eq:BER_u_MF_all}.

Then, we need to calculate the average BER, denoted by $P_{s}^{\prime}$,  for all possible cases. We denote by $P_{1}$ the probability that the signal of desirable user is jammed by an interfering user in MFH communications. Since mode and frequency are two disjoint events, the probability $P_{1}$ is equal to $1/(NQ)$. Thus, we can obtain $P_{s}^{\prime}$ for all possible cases as Eq.~\eqref{eq:BER_P_all_MFH}.

Observing the BER of MFH scheme, we can find that the average SINR, the number of interfering users, the number of OAM-modes, the number of available frequency bands, and the number of hops impact on the BER. Although the BER has complex form in the MFH system, we can also directly obtain some results. First, the BER monotonically increases as the average SINR increases. Hence, increasing the transmit power and mitigating interference can decrease the BER of MFH scheme. Second, the BER monotonically increases as the number of interfering users increases. Also, the BER monotonically decreases as the number of OAM-modes increases. In addition, the BER monotonically decreases as the number of available frequency bands increases. Thus, increasing the number of frequency bands can decrease the BER in the MFH system. Moreover, the BER monotonically decreases as the number of hops increases in the MFH system.

In the MFH system, the complexity of MFH system mainly depends on OAM-transmitter, mode synthesizer, frequency synthesizer, OAM-receiver, integrator, and EGC. The OAM-transmitter and OAM-receiver can be UCA antenna, which consists of several array-elements distributed equidistantly around the perimeter of receive circle. Since OAM signal can be transmitted within one antenna which equips several array-elements while using a single radio frequency chain, OAM-transmitter and OAM-receiver based UCA can be considered as single radio frequency chain antenna. Mode synthesizer and frequency synthesizer select an OAM-mode and a range of frequency, respectively, to de-hop the received OAM signal. Similar to the integrator in the MH system, integrator in the MFH system can also decompose the received OAM signal. EGC reception which has low complexity among existing receiver models, such as maximal ratio combining, EGC, and selection combining, is used for summing up the square of the signal with equal probability, thus obtaining the received instantaneous SINR. The complexity of FH mainly depends on transmit antenna, receive antenna, frequency synthesizer, and EGC. The complexity of EGC used in MFH system is similar to the complexity of EGC used in the FH system. Although the FH system uses single transmit antenna and receive antenna, the MH system uses OAM-transmitter based UCA and OAM-receiver based UCA which can be considered as single radio frequency chain antenna. Therefore, our developed MFH system and the conventional FH system belong to the category of single radio frequency chain. In addition, the MFH system adds a simple integrator and two mode synthesizers. The complexity of mode synthesizer is similar to the complexity of frequency synthesizer.

Since signal can be transmitted within the new mode dimension and the frequency dimension, our developed MFH scheme can be used for achieving better anti-jamming results for various interfering waveforms such as the wideband noise interference, partial-band noise interference, single-tone interference, and multitone interference. Thus, our developed MFH scheme can potentially be applied into various scenarios, such as wireless local area networks, indoor wireless communication, satellite communication, underwater communication, radar, microwave and so on.

\section{Performance Evaluations}\label{sec:nume}

\begin{figure}
\centering
\includegraphics[scale=0.65]{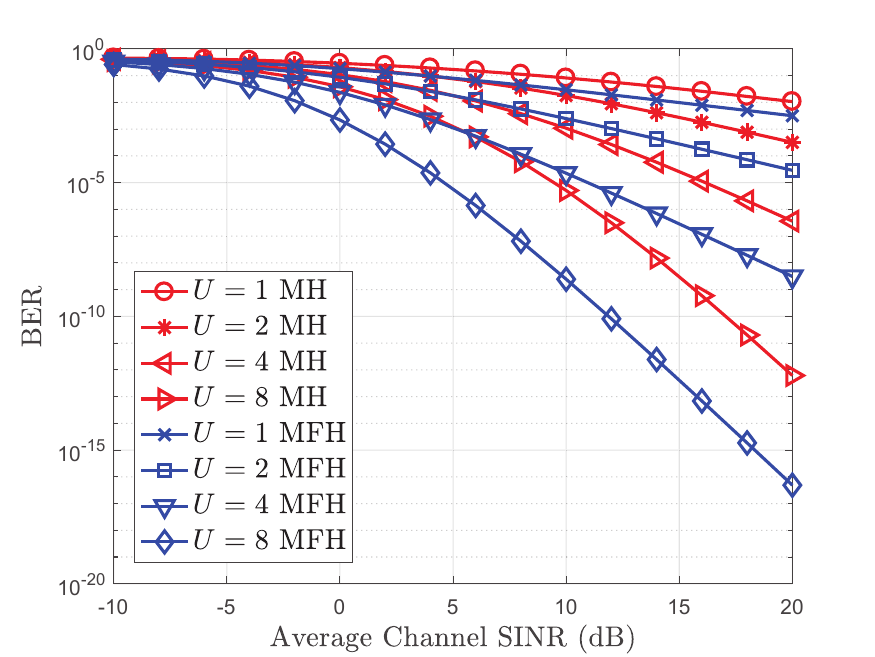}\\
\caption{The BERs of our developed MH and MFH schemes using binary DPSK modulation versus average channel SINR for multiple users scenario, respectively.}
\label{fig:MHMA_MFHMA_SNR}
\end{figure}

In this section, we evaluate the performances for our developed MH and MFH schemes and also compare the BERs of our developed schemes with the conventional wideband FH scheme. Numerical results for anti-jamming evaluated with different mode/frqeucny hops, SINR, and different interfering users are presented. In Figs.~\ref{fig:mode_hopps},~\ref{fig:MHMA_MFHMA_SNR},~\ref{fig:snr_K},~\ref{fig:BER_3D_K_N}, and~\ref{fig:BER_3D_N_q}, we employ binary DPSK modulation to evaluate the BER of the systems. Figs.~\ref{fig:BER_SNR_FSK_DPSK} and~\ref{fig:BER_SNR_MFH_DPSK_FSK} depict the effect of binary DPSK modulation and non-coherent binary FSK modulation on the BER of our developed schemes. 
Throughout the simulations, we set $m$ as 1. The numerical results prove that our developed MH scheme within the narrow frequency band can achieve the same BER as the conventional FH scheme and our developed MFH scheme can achieve the smallest BER among these three schemes.

\subsection{BERs for Single User Scenario }

\begin{figure}
\centering
\includegraphics[scale=0.65]{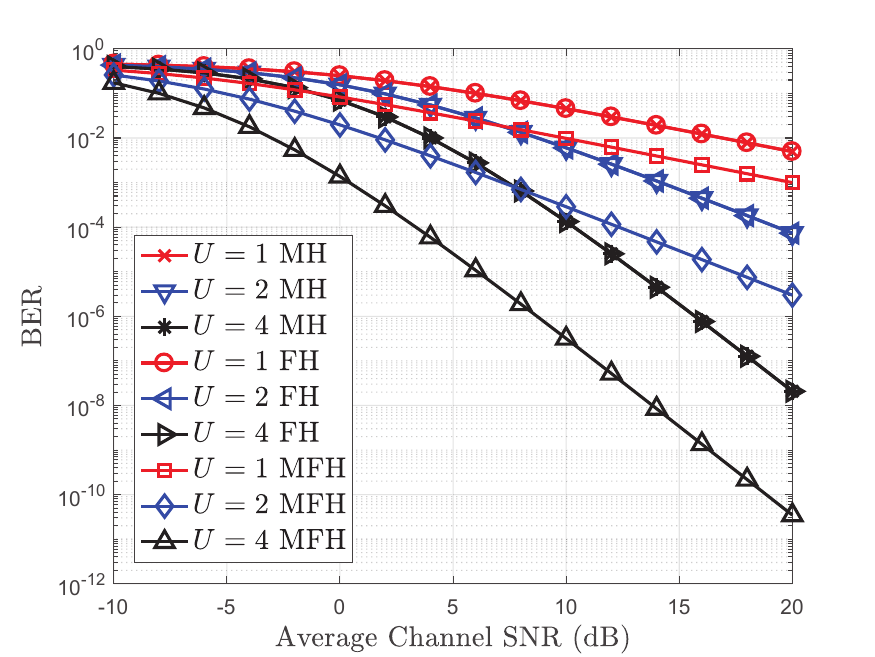}\\
\caption{The BERs of FH, MH, and MFH schemes using binary DPSK modulation versus average channel SNR, respectively.}
\vspace{-15pt}
\label{fig:mode_hopps}
\end{figure}
Figure~\ref{fig:mode_hopps} shows the BERs of FH, MH, and MFH schemes using binary DPSK modulation for single user scenario versus average channel SNR with different hops, where we set $U$ as 1, 2, and 4, respectively. Clearly, the BERs of FH and MH schemes are the same with each other given a fixed hop number. The BER of our developed MFH scheme is the smallest one among the three schemes. The BERs of the three schemes decrease as the average channel SNR increases. Given the hop number, the BERs of our developed MH and MFH schemes are very close to each other in the low SNR region while the differences are widening in the high SNR region. In addition, the BERs of the three schemes decrease as $U$ increases. What's more, the curve of BER falls much faster with higher number of $U$. For example, comparing the BERs with the mode/frequency hops $U= 2$ and $ U = 4$, we can find that the BER of our developed MH scheme for single user scenario decreases from $5.4 \times 10^{-3}$ to $1.3 \times 10^{-4}$ with $U=4$ hops when the average channel SNR increases from 5 dB to 10 dB. Also, the BER decreases from $4.0 \times 10^{-2}$ to $6.0 \times 10^{-3}$ with $U = 2$ hops. As SNR increases from 10 dB to 15 dB, the BER reduces by about $10^{-2}$ with $U = 4$ hops and $10^{-1}$ with $U = 2$ hops. As shown in Fig.~\ref{fig:mode_hopps}, the BER of MH scheme is smaller than that of the scenario with $U=1$. Observing Eqs. (21) and (37), we have $\zeta/(m+\mu \zeta)=1/(\frac{m}{\zeta}+\mu)$ and $G\zeta/(m+G\mu \zeta)=G/(\frac{m}{\zeta}+G\mu)$. Thus, for both MH and MFH schemes the BER decreases as the average SNR increases. Since the received SNR of MFH scheme is $G$ times the received SNR of MH scheme, the BER of MFH scheme is smaller than that of MH scheme, which is consistent with the numerical results. Obtained results verify that the anti-jamming performance of our developed schemes can be better with the higher number of hops and higher SNR. Also, MFH scheme can achieve the best anti-jamming performance among the three schemes.

\subsection{BERs for Multiple Users Scenario}

Figure~\ref{fig:MHMA_MFHMA_SNR} compares the BERs of our developed MH and MFH schemes using DPSK modulation versus average channel
SINR, where we set the number of interfering users as 10, the available number of FH as 5, available number of MH 10, and the number of mode / frequency hops as 1, 2, 4, and 8, respectively. Given fixed number of hops, the BER of our developed MFH scheme is smaller than that of MH scheme in the average channel SINR region for multiple users scenario. Only in low SINR region, the BERs between MFH system and MH system are close to each other. This is because the probability jammed by interfering users in the MFH scheme is smaller than that in the MH scheme. Results prove that our developed MH scheme can be jointly used with the conventional FH scheme to achieve lower BER.

\begin{figure}
\centering
\includegraphics[scale=0.56]{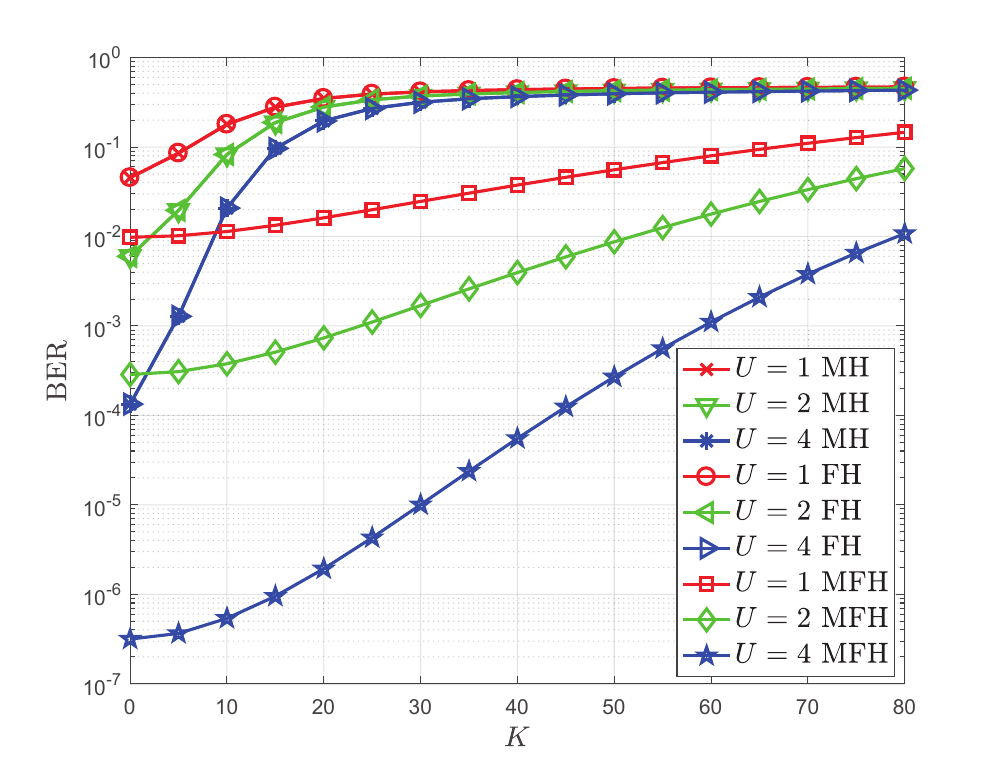}\\
\caption{The BERs versus the number of interfering users with FH, MH, and MFH schemes using binary DPSK modulation, respectively.}
\label{fig:snr_K}
\end{figure}

\begin{figure}
\centering
\includegraphics[scale=0.6]{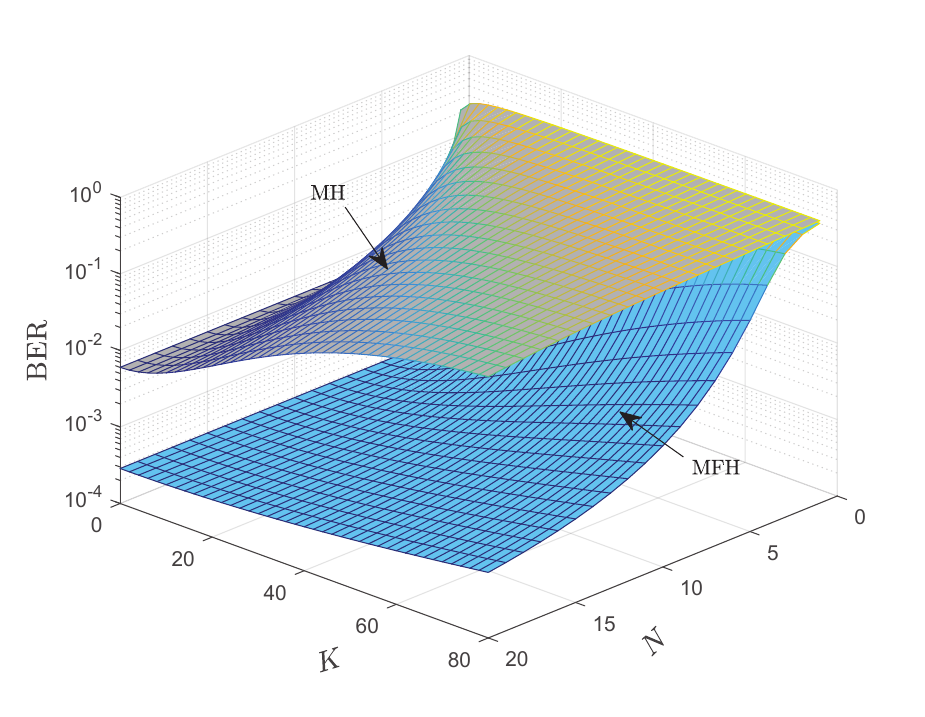}\\
\caption{The BERs versus different number of interfering users and the available number of OAM-modes using our developed MH and MFH schemes using binary DPSK modulation, respectively.}
\vspace{-15pt}
\label{fig:BER_3D_K_N}
\end{figure}

Figure~\ref{fig:snr_K} shows the BERs of FH, MH, and MFH schemes using binary DPSK modulation versus the number of interfering users with different hops for interfering multiple users scenario, where we set average SINR as 10 dB, and $U$ as 1, 2, and 4, respectively.
As shown in Fig.~\ref{fig:snr_K}, the BERs of FH, MH, and MFH schemes for interfering multiple users scenario are in proportion to the number of interfering users. As the number of interfering users increases, the BERs of the three schemes increase until BERs are very close to a fixed value. In addition, comparing BERs of our developed MH scheme and MFH scheme, we can obtain that the BER of MFH scheme is much smaller than that of the MH scheme. Results coincide with the fact that the anti-jamming performance of our developed MH and MFH schemes are better with smaller number of interfering users. Moreover, our developed MFH scheme can significantly improve the anti-jamming performance.

Figure~\ref{fig:BER_3D_K_N} shows the BERs of our developed MH and MFH schemes using binary DPSK modulation versus the number of interfering users and the  number of OAM-modes, where we set SINR as 10 dB, the number of OAM-mode hops as $U = 2$, and the available number of FH as 5. We can obtain that the BERs decrease as the number of OAM-modes increases and the number of interfering users decreases. This is because the number of OAM-modes increases as the probability jammed by interfering users decreases. The number of interfering users decreases as the probability jammed by interfering users decreases, thus increasing the received SNR. Moreover, the BER of our developed MFH scheme is much smaller than the MH scheme. Results prove that the probability of desired signal jammed by interfering user decreases as the number of OAM-modes increases and the number of interfering users decreases, thus guaranteeing the transmission reliability.

\begin{figure}
\centering
\includegraphics[scale=0.6]{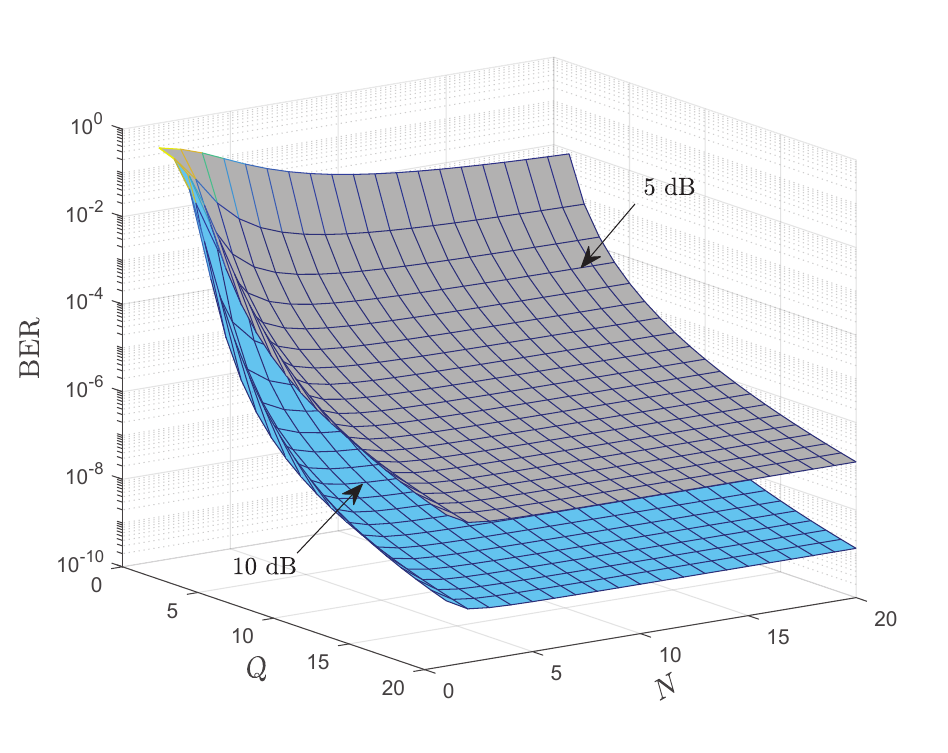}\\
\caption{The BER versus the available number of frequency bands and the available number of OAM-modes with our developed MFH scheme using binary DPSK modulation.}
\label{fig:BER_3D_N_q}
\end{figure}

Figure~\ref{fig:BER_3D_N_q} shows the BERs of our developed MFH scheme using binary DPSK modulation versus the available number of frequency bands and the available number of OAM-modes, where we set the number of interfering users as 10, the number of hops as 4, and the average channel SINR as 5 dB and 10 dB. The BER decreases as the available number of frequency bands and the available number of OAM-modes increase. This is because the probability that the signal of desired user is jammed by interfering users decreases as the number of OAM-modes and frequency bands increase, leading to the downgrading of corresponding BER. Fig.~\ref{fig:BER_3D_N_q} also proves that the BER decreases as the average SINR increases.

\subsection{Comparison Between Binary DPSK and FSK Modulations for MH and MFH Schemes}

\begin{figure}
\centering
\includegraphics[scale=0.65]{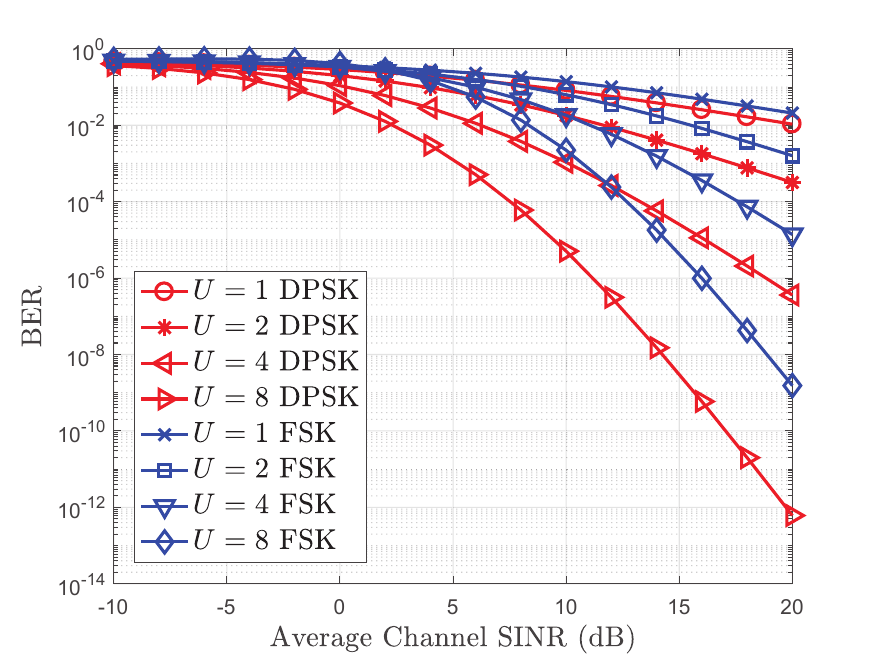}\\
\caption{The BER of our developed MH scheme versus average channel SINR with binary DPSK and FSK modulation, respectively.}
\label{fig:BER_SNR_FSK_DPSK}
\end{figure}

Figure~\ref{fig:BER_SNR_FSK_DPSK} compares the BERs of our developed MH scheme using binary DPSK modulation and binary FSK modulation, for the multiple users scenario versus the average channel SINR, where we set the number of OAM-mode hops as 1, 2, 4, and 8, respectively. The results in Fig.~\ref{fig:BER_SNR_FSK_DPSK} show that the BER is smaller of using binary DPSK modulation than that of using binary FSK modulation. The BERs of using binary DPSK and FSK modulation are close to each other in low SINR region while the BER difference between binary DPSK and FSK modulation increases in the high SINR region. This is because $\zeta/(m+\mu \zeta)$ for MH scheme increases as $\mu$ decreases. Therefore, the BER corresponding to binary DPSK modulation ($\mu=1$) is smaller than that corresponding to binary FSK modulation ($\mu=1/2$). Clearly, the BER of the MH scheme using binary DPSK modulation decreases as the number of hops increases.

\begin{figure}
\centering
\includegraphics[scale=0.65]{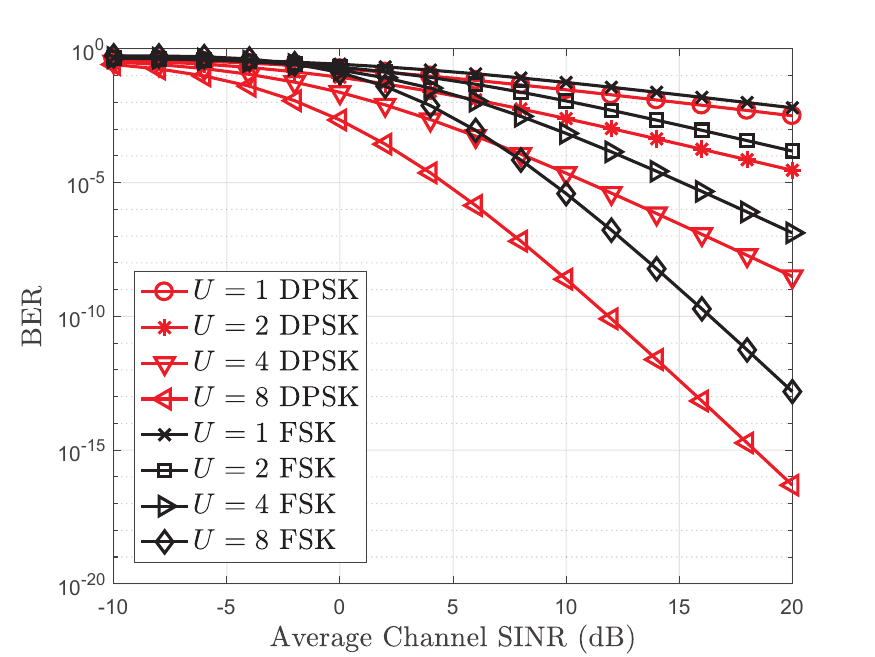}\\
\caption{The BER of our developed MFH scheme versus average channel SINR with binary DPSK and FSK modulation, respectively.}
\label{fig:BER_SNR_MFH_DPSK_FSK}
\end{figure}

Figure~\ref{fig:BER_SNR_MFH_DPSK_FSK} shows the BERs of our developed MFH scheme using binary DPSK and BFSK versus the average channel SINR for multiple users scenario, where we set the number of interfering users as 10, the available number of FH as 2, the available number of MH as 10, the number of OAM-mode hops as 1, 2, 4, and 8, respectively. As shown in Fig.~\ref{fig:BER_SNR_MFH_DPSK_FSK}, we also can obtain that the BER of MFH scheme using binary DPSK modulation is smaller than that using binary FSK modulation. This is because $G\zeta/(m+G\mu \zeta)$ for MFH scheme increases as $\mu$ decreases. Observing Figs.~\ref{fig:BER_SNR_FSK_DPSK} and \ref{fig:BER_SNR_MFH_DPSK_FSK}, the BERs of MFH scheme using binary FSK modulation are larger than that of MH scheme.

\section{Conclusions} \label{sec:conc}
In this paper, we proposed the MH scheme which is expected to be a new technique for anti-jamming in wireless communications. To evaluate the anti-jamming performance, we derived the generic closed-form expression of BER with our developed MH scheme. Since our developed MH scheme provides a new angular/mode dimension within narrow frequency band for wireless communications, the anti-jamming results of our developed
MH scheme can be the same with that of the conventional wideband FH schemes. Furthermore, we proposed MFH scheme to further enhance the anti-jamming performance for wireless communications. We also derived the closed-form expression of BER with our developed MFH scheme to analyze the anti-jamming results. Numerical results show that our developed MH scheme within the narrow frequency has the same anti-jamming
results as compared with the conventional wideband FH schemes and that the BER of our developed MH scheme decreases as the number of hops increases, the number of interfering users decreases, and the average channel SINR increases. Moreover, our developed MFH scheme outperforms
the conventional FH schemes. In addition, our developed MH and MFH schemes using binary DPSK modulation can achieve better anti-jamming results than those using binary FSK modulation.


\setcounter{TempEqCnt}{\value{equation}} 
\setcounter{equation}{46} 
\begin{figure*}[ht]
\begin{eqnarray}
 \Phi_{\gamma_{s}}(w)\hspace{-0.2cm}&=&\hspace{-0.2cm}m^{mU}\prod_{u=1}^{U-V}\left(\frac{1}{m-j w \zeta}\right)^{m}
 \prod_{v=1}^{V}\left(\frac{1}{m-j w \overline{\delta}_{v}}\right)^{m}\nonumber\\
  &=& \hspace{-0.2cm}\left\{\begin{array}{lll}
 \hspace{-0.2cm}m^{mU}\left[\sum\limits_{u_{1}=1}^{m(U-V)}\frac{P_{u_{1}}}{(m-jw \zeta)^{m\left(U-V\right)-u_{1}+1}}
 +\sum\limits_{u_{2}=1}^{ma}\frac{Q_{u_{2}}}{(m-jw \overline{\delta}_{L})^{ma-u_{2}+1}}
    +\sum\limits_{v=1}^{V-a}\sum\limits_{u_{3}=1}^{m}\frac{W_{vu_{3}}}{\left(m-jw\overline{\delta}_{v}\right)^{m-u_{3}+1}}\right],
    a \geq 1 ;
    \vspace{0.1cm}\\
 \hspace{-0.2cm}m^{mU}\left[\sum\limits_{u_{1}=1}^{m(U-V)}\frac{P_{u_{1}}}{(m-jw \zeta)^{m\left(U-V\right)-u_{1}+1}}
    +\sum\limits_{v=1}^{V}\sum\limits_{u_{3}=1}^{m}\frac{W_{vu_{3}}}{\left(m-jw\overline{\delta}_{v}\right)^{m-u_{3}+1}}\right],
     \hspace{4cm} a = 0 , \\
    \end{array}
    \right.
     \label{eq:phi_f}
\end{eqnarray}
\hrulefill
\end{figure*}
\setcounter{equation}{\value{TempEqCnt}}

\begin{appendices}
\section{Proof of Theorem 1}\label{Appendix:_A}
Given $V$ hops jammed by corresponding $D_{v}$ interfering users, the BER, denoted by $P_{b}(\gamma_{s},V, D_{v}|U)$, for $U$ hops with EGC reception at the receiver is given as follows~\cite{Digital_C}:

\setcounter{equation}{39}
\begin{align}
    P_{b}(\gamma_{s},V, D_{v}|U)=2^{1-2U}e^{-\mu\gamma_{s}}\sum_{v_{1}=0}^{U-1}c_{v_{1}}\gamma_{s}^{v_{1}},\nonumber\\
    \label{eq:BER_AGWN}
\end{align}
where $c_{v_{1}}$ is given by
\begin{eqnarray}
    c_{v_{1}}=\frac{1}{v_{1}!}\sum_{v_{2}=0}^{U-v_{1}-1}\dbinom{2U-1}{v_{2}}.
\end{eqnarray}

\setcounter{TempEqCnt}{\value{equation}} 
\setcounter{equation}{48} 
\begin{figure*}[ht]
\begin{small}
\begin{eqnarray}
p_{\gamma_{s}}(\gamma_{s})\hspace{-0.2cm}&=&\hspace{-0.2cm}\frac{1}{2 \pi}\int_{\infty}^{\infty}\Phi_{\gamma_{s}}(w) e^{-jw \gamma_{s}} dw\nonumber\\
  &=& \hspace{-0.2cm}\left\{\begin{array}{lll}
 \hspace{-0.2cm}\frac{m^{mU}}{2 \pi j}\displaystyle\int_{-j\infty}^{+j\infty}\sum\limits_{u_{1}=1}^{m(U-V)}\frac{P_{u_{1}}}{(m-jw \zeta)^{m\left(U-V\right)-u_{1}+1}}e^{-jw \gamma_{s}}d(jw)
 \!\!+\!\!\frac{m^{mU}}{2 \pi j}\displaystyle\int_{-j\infty}^{+j\infty}\sum\limits_{u_{2}=1}^{ma}\frac{Q_{u_{2}}}{(m-jw \overline{\delta}_{L})^{ma-u_{2}+1}}e^{-jw \gamma_{s}}d(jw)\\
    +\frac{m^{mU}}{2 \pi j}\displaystyle\int_{-j\infty}^{+j\infty}\sum\limits_{v=1}^{V-a}\sum\limits_{u_{3}=1}^{m}\frac{W_{vu_{3}}}
    {\left(m-jw\overline{\delta}_{v}\right)^{m-u_{3}+1}}e^{-jw \gamma_{s}}d(jw), \hspace{6.0cm} a \geq 1 ;
    \vspace{0.1cm}\\
 \hspace{-0.2cm}\frac{m^{mU}}{2 \pi j}\!\!\! \displaystyle\int_{-j\infty}^{+j\infty}\sum\limits_{u_{1}=1}^{m(U-V)}\frac{P_{u_{1}}}{(m-jw \zeta)^{m\left(U-V\right)-u_{1}+1}}e^{-jw \gamma_{s}}d(jw)
    \!+\!\frac{m^{mU}}{2 \pi j}\displaystyle\int_{-j\infty}^{+j\infty}\sum\limits_{v=1}^{V}\!\!\sum\limits_{u_{3}=1}^{m}
    \frac{W_{vu_{3}}}{\left(m-jw\overline{\delta}_{v}\right)^{m-u_{3}+1}}e^{-jw \gamma_{s}}d(jw),\\
    \hspace{14.1cm}  a = 0 ,
    \end{array}
    \right.
     \label{eq:p_gamma_s_generic}
\end{eqnarray}
\end{small}
\hrulefill
\end{figure*}
\setcounter{equation}{\value{TempEqCnt}}

For Nakagami-$m$ fading, the probability density function (PDF), denoted by $p_{\gamma}(\gamma)$, can be expressed as follows:
\begin{eqnarray}
p_{\gamma}(\gamma)=\frac{\gamma^{m-1}}{\Gamma(m)}\left(\frac{m}{\overline{\gamma}}\right)^{m}e^{-m\frac{\gamma}{\overline{\gamma}}},
\end{eqnarray}
where $m$ is the fading parameter, $\Gamma(\cdot)$ is Gamma function, $\gamma$ represents the SINR of channel, and $\overline{\gamma}$ represents the average SINR of channel. Then, $P_{e}(V, D_{v}|U)$ can be expressed as follows:
\begin{eqnarray}
\hspace{-0.4cm} P_{e}(V, D_{v}|U)
 \hspace{-0.3cm}&=&\hspace{-0.4cm}\underbrace{\int_{0}^{\infty}\!\!  \int_{0}^{\infty}\!\! \cdots\!\! \int_{0}^{\infty}}_{U-fold}  P_{b}(\gamma_{s},V,D_{v}|U)\nonumber\\
  &&\hspace{-0.4cm}\left[\prod_{u=1}^{U-V}\!p_{\gamma_{u}}\!(\gamma_{u}\!)\!\!\prod_{v=1}^{V}\!p_{\delta_{v}}\!(\delta_{v})\right]
  \!\!\prod_{u=1}^{U-V}d \!\gamma_{u} \!\!\prod_{v=1}^{V} \!d \delta_{v},
 \label{eq:P_e_gamma}
\end{eqnarray}
which is an $U$-fold integral. For convenience to calculate $P_{e}(V, D_{v}|U)$, we can express the right hand of  Eq.~\eqref{eq:P_e_gamma} into an integral. Thus, $P_{e}(V, D_{v}|U)$ can be expressed as follows:
\begin{eqnarray}
 P_{e}(V, D_{v}|U)=\int_{0}^{\infty}P_{b}(\gamma_{s},V, D_{v}|U)p_{\gamma_{s}}(\gamma_{s})d \gamma_{s},\label{eq:P_e_v_U_MH}
\end{eqnarray}
where $p_{\gamma_{s}}(\gamma_{s})$ is the PDF of $\gamma_{s}$.
Based on the joint PDF, namely $\left[\prod_{u=1}^{U-V}p_{\gamma_{u}}(\gamma_{u})\prod_{v=1}^{V}p_{\delta_{v}}(\delta_{v})\right]$ in Eq.~\eqref{eq:P_e_gamma}, we need to derive the expression of $p_{\gamma_{s}}(\gamma_{s})$.

Based on the PDFs of $\gamma_{u}$ and $\delta_{v}$, we can calculate the characteristic functions, denoted by $\Phi_{\gamma_{u}}(w)$ and $\Phi_{\gamma_{u}}(w)$, for $\gamma_{u}$ and $\delta_{v}$, respectively, with the Fourier transform as follows:
\begin{subnumcases}
    {}\Phi_{\gamma_{u}}(w)=\left(\frac{m}{m-j w \zeta}\right)^{m};
    \\
    \Phi_{\delta_{v}}(w)=\left(\frac{m}{m-j w \overline{\delta}_{v}}\right)^{m}.
\end{subnumcases}

Since instantaneous SINR experiences Nakagami-$m$ fading and fadings on the $U$ channels are mutually statistically independent, instantaneous SINRs are statistically independent. Hence, the characteristic function, denoted by $\Phi_{\gamma_{s}}(w)$, for $\gamma_{s}$ in MH communications can be obtained as follows:
\begin{eqnarray}
    \hspace{-0.3cm}\Phi_{\gamma_{s}}(w)\hspace{-0.2cm}&=&\hspace{-0.2cm} \left[\Phi_{\gamma_{u}}(w)\right]^{U-V}\prod_{v=1}^{V}\Phi_{\delta_{v}}(w)\nonumber\\
    \hspace{-0.2cm}&=&\hspace{-0.2cm}\prod_{u=1}^{U-V}\left(\frac{m}{m-j w \zeta}\right)^{m}\prod_{v=1}^{V}\left(\frac{m}{m-j w \overline{\delta}_{v}}\right)^{m}.
    \label{eq:phi_MU}
\end{eqnarray}

Then, by using partial fraction decomposition algorithm, the Eq.~\eqref{eq:phi_MU} can be re-written as Eq.~\eqref{eq:phi_f}, where $P_{u_{1}}$, $Q_{u_{2}}$, and $W_{vu_{3}}$ are given as follows:
\setcounter{TempEqCnt}{\value{equation}} 
\setcounter{equation}{54} 
\begin{figure*}[ht]
\begin{eqnarray}
p_{\gamma_{s}}(\gamma_{s})\!\!=\!\! \left\{
\begin{array}{lll}
\hspace{-0.2cm}m^{mU}\left\{\sum\limits_{u_{1}=1}^{m\left(U-V\right)}\frac{P_{u_{1}}e^{-\frac{m}{\zeta}\gamma_{s}}
\gamma_{s}^{m(U-V)-u_{1}}}{\zeta^{m(U-V)-u_{1}+1}\Gamma[m(U-V)-u_{1}+1]}
+\sum\limits_{u_{2}=1}^{ma}\frac{Q_{u_{2}}e^{-\frac{m}{\overline{\delta}_{L}}\gamma_{s}}
\gamma_{s}^{ma-u_{2}}}{\overline{\delta}_{L}^{ma-u_{2}+1}\Gamma[ma-u_{2}+1]}
+\sum\limits_{v=1}^{V-a}\sum\limits_{u_{3}=1}^{m}\frac{W_{vu_{3}}e^{-\frac{m}{\overline{\delta}_{v}}\gamma_{s}}
\gamma_{s}^{m-V}}{\overline{\delta}_{v}^{m-V+1}\Gamma[m-V+1]}
\right\}, \\
\hspace{14.5cm}a \geq 1;
\\
\hspace{-0.2cm}m^{mU}\left\{\sum\limits_{u_{1}=1}^{m\left(U-V\right)}\frac{P_{u_{1}}e^{-\frac{m}{\zeta}\gamma_{s}}
\gamma_{s}^{m(U-V)-u_{1}}}{\zeta^{m(U-V)-u_{1}+1}\Gamma[m(U-V)-u_{1}+1]}
+\sum\limits_{v=1}^{V}\sum\limits_{u_{3}=1}^{m}\frac{W_{vu_{3}}e^{-\frac{m}{\overline{\delta}_{v}}\gamma_{s}}
\gamma_{s}^{m-V}}{\overline{\delta}_{v}^{m-V+1}\Gamma[m-V+1]}
\right\}, \hspace{3.3cm} a=0.
\\
\end{array}
\right.
\label{eq:p_gamma_s}
\end{eqnarray}
\hrulefill
\end{figure*}
\setcounter{equation}{\value{TempEqCnt}}

\setcounter{equation}{47}
\begin{subnumcases}
    {}\!\!\!\!P_{u_{1}}\!\!=\!\frac{1}{(\!u_{1}-\!\!1)!} \frac{d^{u_{1}\!-\!1}}{d(jw)^{\!u_{1}\!-\!1}}
   \!\! \left[\!(m\!\!-\!\!jw\zeta\!)^{m(U\!-\!V)}\!\Phi_{\gamma_{s}}\!(w)\!\right]\!\!\Bigg|_{jw=\frac{m}{\zeta}};\nonumber\\
   \\
   \!\!\!\! Q_{u_{2}}\!\!=\!\!\frac{1}{(u_{2}\!\!-\!\!1)!} \frac{d^{u_{2}-1}}{d(j w)^{u_{2}-1}}\!\!\left[(m\!\!-\!\!jw\overline{\delta}_{L})^{ma}\Phi_{\gamma_{s}}(w)\right]\Bigg|_{jw=\frac{m}{\overline{\delta}_{L}}};\nonumber\\
   \\
   \!\!\!\! W_{vu_{3}}=\frac{1}{(\!u_{3}-\!1)!} \frac{d^{u_{3}\!-\!1}}{d(jw)^{\!u_{3}\!-\!1}} \left[\!(m\!\!-\!\!jw\overline{\delta}_{v}\!)^{m}\!\Phi_{\gamma_{s}}\!(w)\!\right]\!\!\Bigg|_{jw=\frac{m}{\overline{\delta}_{v}}}.\nonumber\\
\end{subnumcases}
Thus, we can obtain the generic PDF of $\gamma_{s}$ as Eq.~\eqref{eq:p_gamma_s_generic}. When $a \geq 1$, the first term on the right hand of Eq.~\eqref{eq:p_gamma_s_generic} can be derived as follows:
\setcounter{equation}{49}
\begin{eqnarray}
    &&\hspace{-0.75cm}\frac{m^{mU}}{2 \pi j}\displaystyle\int_{-j\infty}^{+j\infty}\sum\limits_{u_{1}=1}^{m(U-V)}\!\!\!\frac{P_{u_{1}}}{(m-jw \zeta)^{m\left(U-V\right)-u_{1}+1}}e^{-jw \gamma_{s}}d(jw)\nonumber\\
    &&\hspace{-0.75cm}=\sum\limits_{u_{1}=1}^{m(U-V)}\!\!\!\frac{m^{mU} P_{u_{1} e^{-m\gamma_{s}/\zeta } }}{\zeta^{m\left(U-V\right)-u_{1}+1} \times 2\pi j} \!\!\displaystyle\int_{-j\infty+\frac{m}{\zeta}}^{+j\infty \frac{m}{\zeta}} \frac{e^{\gamma_{s} z}}{z^{m\left(U-V\right)-u_{1}+1}} dz. \nonumber\\
    \label{eq:P_u_1}
\end{eqnarray}
In the integral part, $1/z$ approaches to 0 as $Re(z)$ approaches to $\infty$. Using Cauchy's theorem and Residue theorem~\cite{2010_Fourier}, we can obtain
\begin{equation}
    \frac{1}{2\pi j}\!\!\int_{-j \infty +\frac{m}{\zeta}}^{+j \infty +\frac{m}{\zeta}}\!\! \frac{e^{\gamma_{s}z} }{z^{m\left(U-V\right)-u_{1}+1}} d z \!=\!\frac{1}{2\pi j}\!\!\int_{C} \!\frac{e^{\gamma_{s}z} }{z^{m\left(U-V\right)-u_{1}+1}} d z, \\
\end{equation}
where $C$ is an open contour from initial point to negative infinity on the real axis. According to the characteristics of Gamma function~\cite{1999_functions}, we have
\begin{eqnarray}
  \frac{1}{2\pi j}\int_{C} \frac{e^{\gamma_{s}z} }{z^{m\left(U-V\right)-u_{1}+1}} d z=\frac{\gamma_{s}^{^{m(U-V)-u_{1}}}}{\Gamma[m(U-V)-u_{1}+1]},
\end{eqnarray}
where $\Gamma(\cdot)$ is the Gamma function and given as follows:
\begin{eqnarray}
    \Gamma[m(U-V)-u_{1}+1]=[m(U-V)-u_{1}+1]! .
\end{eqnarray}
Thus, Eq.~\eqref{eq:P_u_1} can be re-written as follows:
\begin{eqnarray}
    &&\hspace{-0.75cm}\frac{m^{mU}}{2 \pi j}\displaystyle\int_{-j\infty}^{+j\infty}\sum\limits_{u_{1}=1}^{m(U-V)}\frac{P_{u_{1}}}{(m-jw \zeta)^{m\left(U-V\right)-u_{1}+1}}e^{-jw \gamma_{s}}d(jw)\nonumber\\
    &&\hspace{-0.75cm}=\sum\limits_{u_{1}=1}^{m(U-V)}\frac{m^{mU} P_{u_{1} e^{-m\gamma_{s}/\zeta } }}{\zeta^{m\left(U-V\right)-u_{1}+1}}\frac{\gamma_{s}^{^{m(U-V)-u_{1}}}}{\Gamma[m(U-V)-u_{1}+1]}.
\end{eqnarray}
Similar to the analysis above, the other terms on the right hand of Eq.~\eqref{eq:p_gamma_s_generic} also can be derived. Thus, the generic PDF of $\gamma_{s}$ corresponding to Eq.~\eqref{eq:p_gamma_s_generic} can be re-written as Eq.~\eqref{eq:p_gamma_s}.
Then, substituting Eqs.~\eqref{eq:BER_AGWN} and ~\eqref{eq:p_gamma_s} into Eq.~\eqref{eq:P_e_v_U_MH}, we can obtain the $P_{e}(V, D_{v}|U)$ as Eq. ~\eqref{eq:BER_u_MH_all}.
\section{Proof of Theorem 2}\label{Appendix:_C}
For the scenario with $l_{u} \neq l_{u,k}$, the characteristic function of $\gamma_{s}$ corresponding to $U$ hops can be re-written as follows:
\setcounter{equation}{55}
\begin{eqnarray}
  \Phi_{\gamma_{s}}(w)=\left(\frac{m}{m-jw \zeta}\right)^{mU}.
\end{eqnarray}
Thus, the corresponding PDF with Nakagami-$m$ fading for $\gamma_{s}$ can be re-written as follows:
\begin{eqnarray}
 p_{\gamma_{s}}(\gamma_{s})=\left(\frac{m}{\zeta}\right)^{mU}\frac{\gamma_{s}^{mU-1}}{\Gamma(mU)}e^{-\frac{m\gamma_{s}}{\zeta}}.
\end{eqnarray}
Then, $P_{e}(U)$ for the scenario with $l_{u} \neq l_{u,k}$ can be obtained as Eq.~\eqref{eq:BER_P_e_U}.
\end{appendices}

\bibliographystyle{IEEEbib}
\bibliography{References}

\begin{IEEEbiography}[{\includegraphics[width=1in,height=1.25in,clip,keepaspectratio]{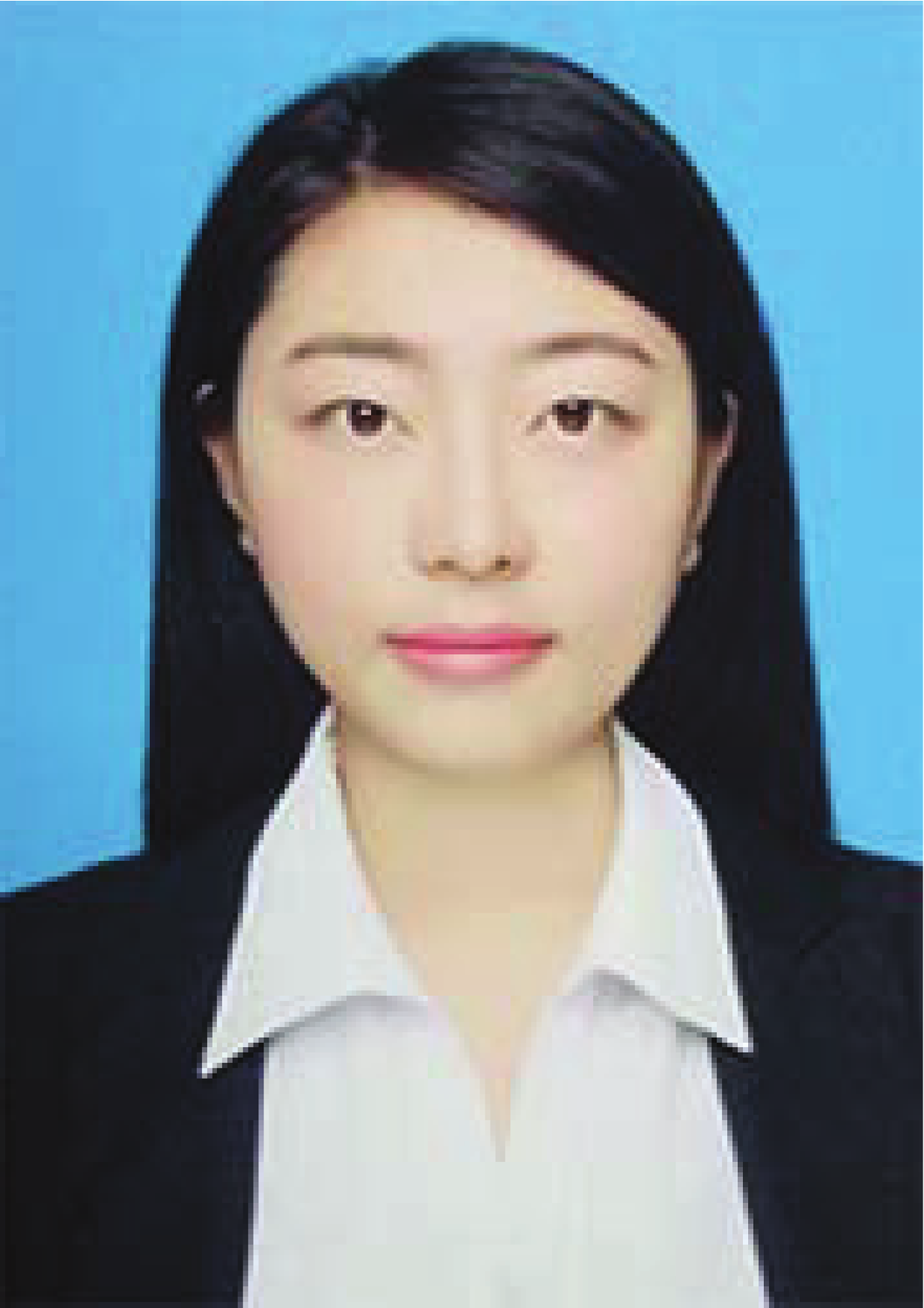}}]{Liping Liang} received B.S. degree in Electronic and Information Engineering from Jilin University, Changchun, China in 2015. She is currently working towards the Ph.D. degree in communication and information system from Xidian University, Xi'an, China. Her research interests focus on 5G wireless communications with emphasis on radio vortex wireless communications and anti-jamming communications.
\end{IEEEbiography}

\begin{IEEEbiography}[{\includegraphics[width=1in,height=1.25in,clip,keepaspectratio]{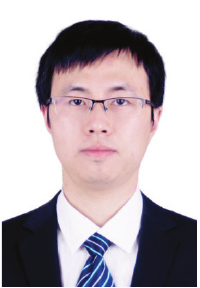}}]{Wenchi Cheng} (M'14) received B.S. degree and Ph.D. degree in Telecommunication Engineering from Xidian University, China, in 2008 and 2014, respectively, where he is an Associate Professor. He joined Department of Telecommunication Engineering, Xidian University, in 2013, as an Assistant Professor. He worked as a visiting scholar at Networking and Information Systems Laboratory, Department of Electrical and Computer Engineering, Texas A\&M University, College Station, Texas, USA, from 2010 to 2011. His current research interests include 5G wireless networks and orbital-angular-momentum based wireless communications. He has published more than 50 international journal and conference papers in IEEE Journal on Selected Areas in Communications, IEEE Magazines, IEEE Transactions, IEEE INFOCOM, GLOBECOM, and ICC, etc. He received the Young Elite Scientist Award of CAST, the Best Dissertation (Rank 1) of China Institute of Communications, the Best Paper Nomination for IEEE GLOBECOM 2014, and the Outstanding Contribution Award for Xidian University. He has served or serving as the Associate Editor for IEEE Access, the Publicity Chair of ICC 2019, the TPC member for IEEE INFOCOM, GLOBECOM, and ICC.
\end{IEEEbiography}

\begin{IEEEbiography}[{\includegraphics[width=1in,height=1.25in,clip,keepaspectratio]{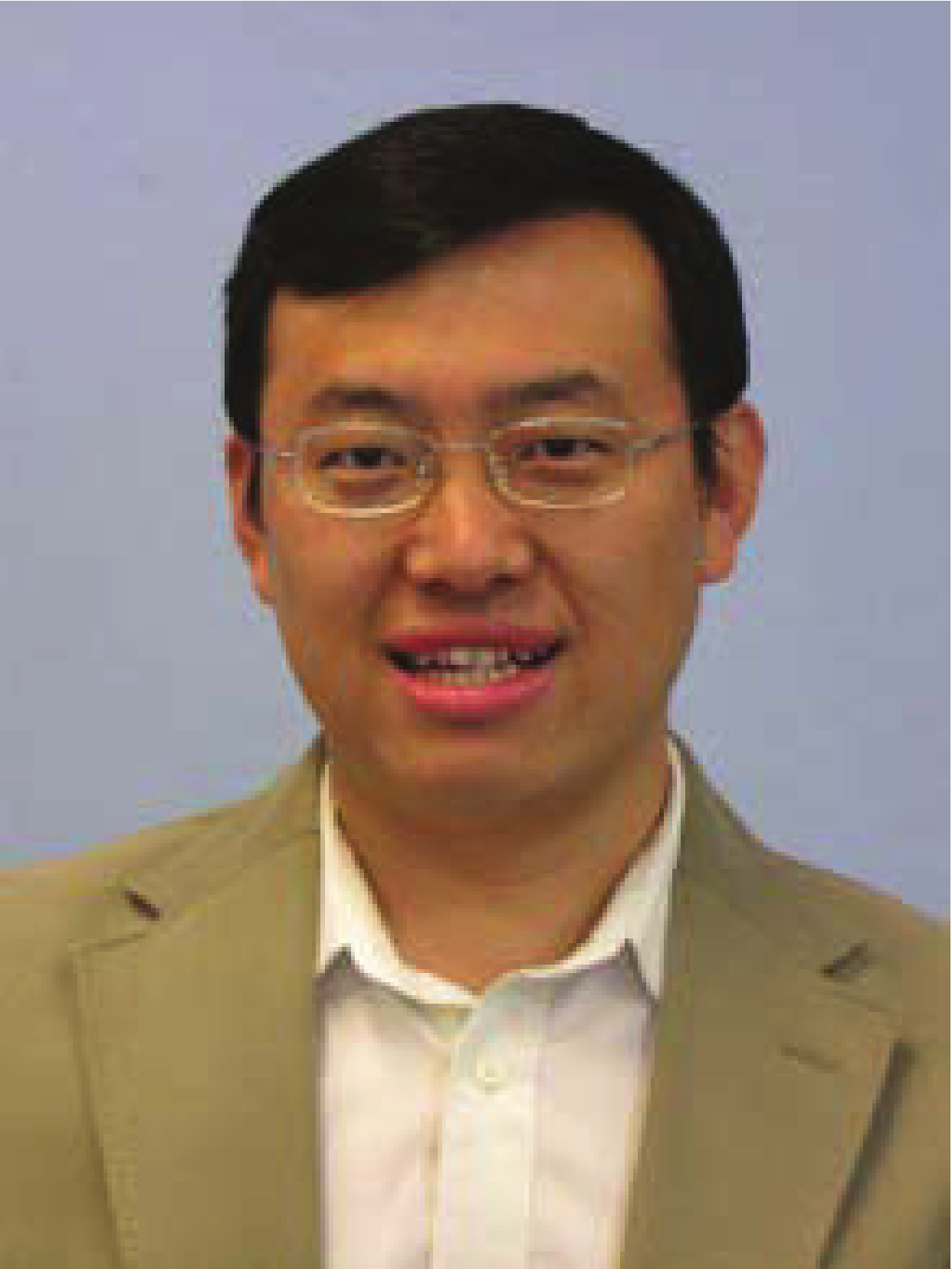}}]{Wei Zhang} (S'01-M'06-SM'11-F'15) received the Ph.D. degree in electronic engineering from the Chinese University of Hong Kong, Hong Kong, in 2005. He was a Research Fellow with the Hong Kong University of Science and Technology, Clear Water Bay, Hong Kong, from 2006 to 2007. He is currently a Professor with the University of New South Wales, Sydney, NSW, Australia. His research interests include cognitive radio, 5G, heterogeneous networks, and massive multiple input and multiple output. He is the Editor-in-Chief of the IEEE WIRELESS COMMUNICATIONS LETTERS. He is also an Editor for the IEEE TRANSACTIONS ON COMMUNICATIONS. He served as an Editor for the IEEE TRANSACTIONS ON WIRELESS COMMUNICATIONS from 2010 to 2015 and the IEEE JOURNAL ON SELECTED AREAS IN COMMUNICATIONS (Cognitive Radio Series) from 2012 to 2014.

He participates actively in committees and conference organization for the IEEE Communications Society. He is Vice Chair for the IEEE Wireless Communications Technical Committee. He is an elected member of the SPCOM Technical Committee of the IEEE Signal Processing Society. He also served in the organizing committee of the 2016 IEEE International Conference on Acoustics, Speech and Signal Processing, Shanghai, China, and the IEEE GLOBECOM 2017, Singapore. He is TPC co-Chair of the 2017 Asia-Pacific Conference on Communications and the 2019 International Conference on Communications in China. He is a member of the Board of Governors of the IEEE Communications Society. He was a recipient of the IEEE Communications Society Asia-Pacific Outstanding Young Researcher Award 2009 and the IEEE ComSoc TCCN Publication Award 2017 as well as four best paper awards in international Conferences.
\end{IEEEbiography}

\begin{IEEEbiography}[{\includegraphics[width=1in,height=1.25in,clip,keepaspectratio]{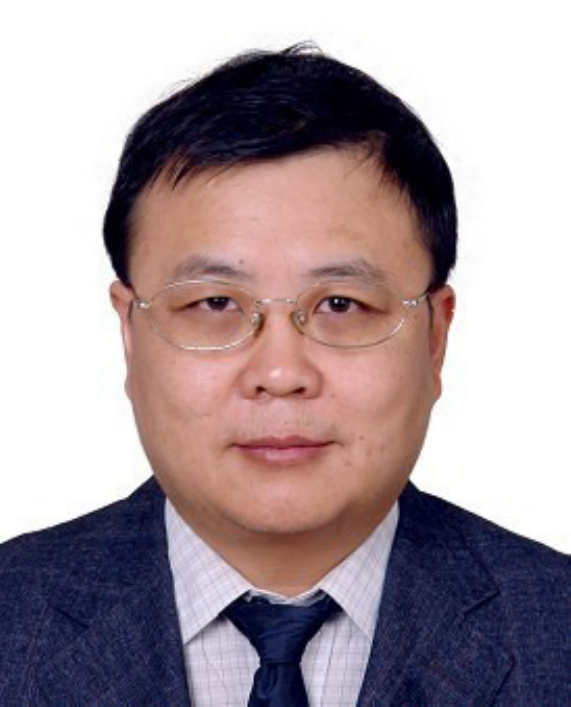}}]{Hailin Zhang} (M'97) received B.S. and M.S. degrees from Northwestern Polytechnic University, Xi'an, China, in 1985 and 1988 respectively, and the Ph.D. from Xidian University, Xi'an, China, in 1991. In 1991, he joined School of Telecommunications Engineering, Xidian University, where he is a senior Professor and the Dean of this school. He is also currently the Director of Key Laboratory in Wireless Communications Sponsored by China Ministry of Information Technology, a key member of State Key Laboratory of Integrated Services Networks, one of the state government specially compensated scientists and engineers, a field leader in Telecommunications and Information Systems in Xidian University, an Associate Director of National 111 Project. Dr. Zhang's current research interests include key transmission technologies and standards on broadband wireless communications for 5G and 5G-beyond wireless access systems. He has published more than 150 papers in journals and conferences.
\end{IEEEbiography}

\end{document}